\documentclass{elsart}

\usepackage[square,comma]{natbib}
\usepackage{graphicx}
\usepackage{pxfonts}
\usepackage{lineno}

\usepackage{amssymb}

\journal{}

\begin{document}
\thispagestyle{empty}
\begin{Large}
\textbf{DEUTSCHES ELEKTRONEN-SYNCHROTRON}

\textbf{\large{Ein Forschungszentrum der Helmholtz-Gemeinschaft}\\}
\end{Large}

DESY 11-096

June 2011

\begin{eqnarray}
\nonumber &&\cr \nonumber && \cr \nonumber &&\cr
\end{eqnarray}
\begin{eqnarray}
\nonumber
\end{eqnarray}
\begin{center}
\begin{Large}
\textbf{Circular polarization control for the European XFEL in the
soft X-ray regime}
\end{Large}
\begin{eqnarray}
\nonumber &&\cr \nonumber && \cr
\end{eqnarray}

\begin{large}
Gianluca Geloni,
\end{large}
\textsl{\\European XFEL GmbH, Hamburg}
\begin{large}

Vitali Kocharyan and Evgeni Saldin
\end{large}
\textsl{\\Deutsches Elektronen-Synchrotron DESY, Hamburg}
\begin{eqnarray}
\nonumber
\end{eqnarray}
\begin{eqnarray}
\nonumber
\end{eqnarray}
ISSN 0418-9833
\begin{eqnarray}
\nonumber
\end{eqnarray}
\begin{large}
\textbf{NOTKESTRASSE 85 - 22607 HAMBURG}
\end{large}
\end{center}
\clearpage
\newpage

\begin{frontmatter}



\title{Circular polarization control for the European XFEL in the soft X-ray regime}


\author[XFEL]{Gianluca Geloni\thanksref{corr},}
\thanks[corr]{Corresponding Author. E-mail address: gianluca.geloni@xfel.eu}
\author[DESY]{Vitali Kocharyan}
\author[DESY]{and Evgeni Saldin}

\address[XFEL]{European XFEL GmbH, Hamburg, Germany}
\address[DESY]{Deutsches Elektronen-Synchrotron (DESY), Hamburg,
Germany}

\begin{abstract}
The possibility of producing  X-ray radiation with high degree of
circular polarization is an important asset at XFEL facilities.
Polarization control is most important in the soft X-ray region.
However, the baseline of the European XFEL, including the soft X-ray
SASE3 line, foresees planar undulators only, yielding
linearly-polarized radiation. It is clear that the lowest-risk
strategy for implementing polarization control at SASE3 involves
adding an APPLE II - type undulator at the end of the planar
undulator, in order to exploit the micro bunching from the baseline
FEL. Detailed experience is available in synchrotron radiation
laboratories concerning the manufacturing of 5 m - long APPLE II
undulators. However, the choice of a short helical radiator leads to
the problem of background suppression.  The driving idea of our
proposal is that the background radiation can be suppressed by
spatial filtering. This operation can be performed by inserting
slits behind the APPLE II radiator, where the linearly-polarized
radiation spot size is about $30$ times larger than the radiation
spot size from the helical radiator. The last 7 cells of the SASE3
undulator are left with an open gap in order to provide a total $42$
m drift section for electron beam and radiation.  Due to the
presence of the drift the linearly-polarized radiation spot size
increases, and the linearly polarized background radiation can be
suppressed by the slits. At the same time, the microbunch structure
is easily preserved, so that intense (100 GW) coherent radiation is
emitted in the helical radiator. We propose a filtering setup
consisting of a pair of water cooled slits for X-ray beam filtering
and of a $5$ m-long magnetic chicane, which creates an offset for
slit installation immediately behind the helical radiator. Electrons
and X-rays are separated before the slits by the magnetic chicane,
so that the electron beam can pass by the filtering setup without
perturbations. Based on start-to-end simulations we present complete
calculations from the SASE3 undulator entrance up to the radiator
exit including the modulated electron beam transport by the FODO
focusing system in the low charge (20 pC) mode of operation.

\end{abstract}

%
%
%
\end{frontmatter}



\section{\label{sec:uno}  Introduction}

The European XFEL \cite{tdr-2006}, which is currently under
construction, will be equipped with planar undulators providing
linearly polarized radiation in the horizontal plane. While it will
be possible to change the photon energy by opening or closing the
gap of the undulators at different operation energies, polarization
control remains problematic. However, circularly polarized X-ray
radiation is an important tool for investigating magnetic materials
and other material science issues, especially in the soft X-ray
region. In particular, the spectral range between $550$ eV and $900$
eV covers absorption edges of transition metals which are of
relevance for technological applications: Cr, Mn, Fe, Co, and Ni.
This spectral region can be covered by the SASE3 undulator in the
fundamental harmonic at the nominal electron beam energy of $14$
GeV.

The possibility of generating variably polarized radiation at SASE3
was recently investigated at the European XFEL and DESY. Several
optimized schemes where considered \cite{YLI1}-\cite{SCH1}.  All
these schemes exploit the microbunching generated in the planar
undulator, and make use of a short helical radiator at the end of
the SASE3 undulator. However, the exploitation of the microbunching
from the planar undulator leads to background problems, since the
linearly-polarized radiation from the planar undulator should be
suppressed.

In order to solve the background issue it has been proposed
\cite{SCH1} that the radiation in the helical radiator can be tuned
to the second harmonic and be therefore characterized by a different
frequency compared to the linearly polarized radiation. However, for
the European XFEL this option cannot cover the most interesting
region between $550$ eV and $900$ eV at the nominal electron beam
energy of $14$ GeV. It will cover this wavelength range only at $10$
GeV, but in this case the hard X-ray lines SASE1 and SASE2 will not
be able to operate at Angstrom wavelength range. Another drawback of
using second harmonic helical radiators is that in order to achieve
a spectral separation of radiation at the first and at the second
harmonic, an extra dispersive X-ray  optics element needs to be
introduced in the photon beam line. Such type of element has not yet
come into operation.

Another possible solution to the background problem has been
proposed in \cite{YLI1}-\cite{YLI3}. In this option, the electron
beam is deflected by a bending system after the planar undulator and
subsequently passes through the helical radiator. In this way the
linearly polarized radiation is separated from the circularly
polarized one, and the polarization properties of the radiator are
completely decoupled from those of the light produced in the planar
undulator. However, the electron beam microbunching must be
preserved through the bending system on the scale of the soft X-ray
radiation wavelength produced in the planar undulator, and this
constitutes a challenge, which began to be addressed in literature
only very recently \cite{YLI3}.

In this paper we propose a third option, which makes use of an
in-line setup. After the passage through the planar undulator, the
electron beam is sent through a straight section drift, and
subsequently through a short APPLE II radiator tuned at the
fundamental harmonic. The background radiation from the planar
undulator is suppressed by a spatial filtering setup consisting of a
water cooled slits pair for X-ray beam filtering and of a $5$ m-long
magnetic chicane, which creates an offset for slit installation
immediately behind the helical radiator. Electrons and X-rays are
separated before the slits by the magnetic chicane, so that the
electron beam can pass by the filtering setup without perturbations.
Compared to the method proposed in \cite{OURC} for the LCLS, the
introduction of the chicane allows to ease the requirements on the
thin Berillium slits in \cite{OURC}, and allows for the installation
of water-cooled power slits of any material to cope with the
increased heat loading due to the high repetition rate of the
European XFEL. However, it should be noted that the method is
applicable at other facilities, including as well the LCLS. The
helical radiator is placed immediately behind the SASE3 undulator,
consisting of $21$ cells. The last 7 cells of the SASE3 undulator
are left with an open gap in order to provide a total $42$ m drift
section for electron beam and radiation.  Due to the presence of the
drift, the linearly-polarized radiation spot size at the slits is
about $30$ times larger than the circularly-polarized one, and the
linearly polarized background can be easily suppressed by the slits.
At the same time, the in-line design allows to preserve the
microbunching structure. Intense coherent radiation in the $100$ GW
level is emitted in the helical radiator.

\begin{figure}[tb]
\begin{center}
\includegraphics[width=0.75\textwidth]{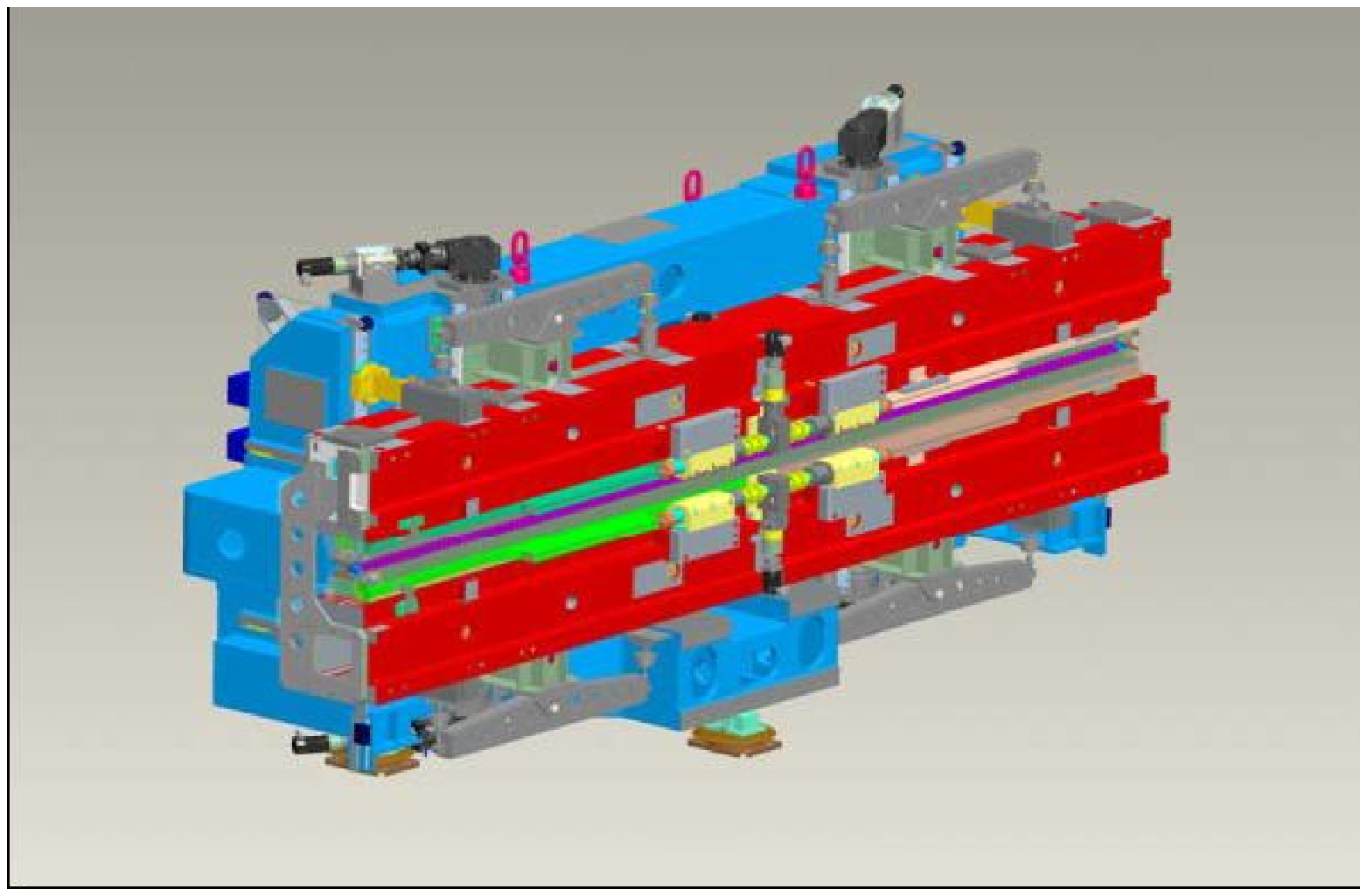}
\end{center}
\caption{Mechanical layout of an APPLE II undulator module (Courtesy
of M. Tischer).} \label{AppleII}
\end{figure}
Simulations performed with the code GENESIS show that in order to
transport the microbunched electron beam through the 42 m-long
straight section (corresponding to the 7 undulator modules with open
gap), it is sufficient to use the existing undulator FODO structure
with the usual (15 m) beta function as a focusing system. The
proposed option has advantages in terms of cost and time, also
because we can afford to use the existing design of APPLE II type
undulators, Fig. \ref{AppleII}, improved for PETRA III \cite{BAHR}.

\section{\label{sec:due} Scheme for circular polarization control based on spatial filtering
of the linearly-polarized radiation background behind the helical
radiator }

\begin{figure}[tb]
\includegraphics[width=1.0\textwidth]{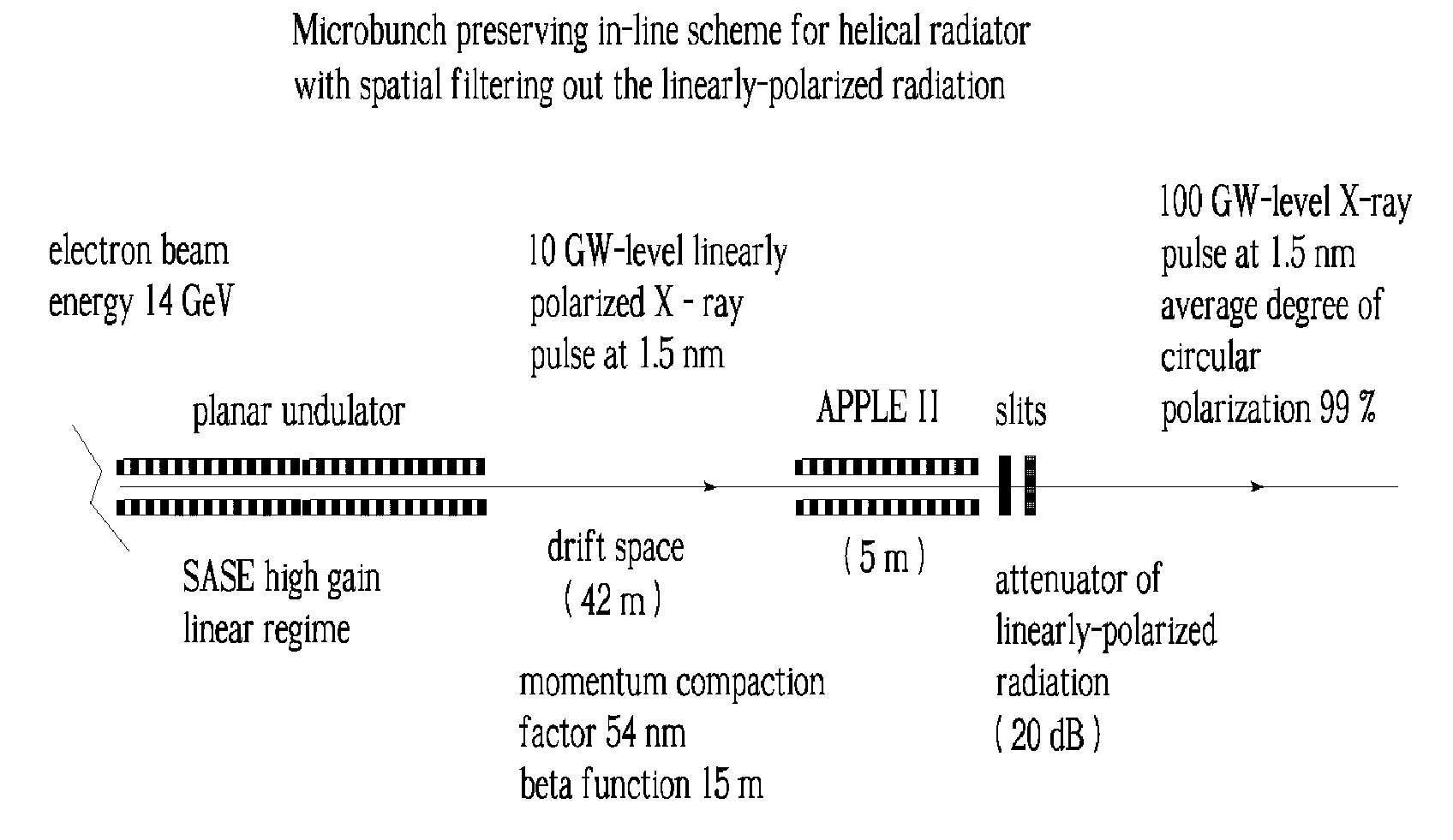}
\caption{Concept for circular polarization control at the European
XFEL. After the planar SASE3 undulator line, the electron beam is
propagated along a $42$ m-long (7 cells of SASE3) straight section.
Subsequently, it passes through a helical radiator. The microbunch
is preserved, and intense coherent radiation is emitted in the
helical radiator. Linearly-polarized radiation  from the planar
undulator is easily suppressed by spatial  filtering with the help
of slits downstream of the helical radiator,  which do not attenuate
the circularly-polarized radiation from the helical radiator. The
electron beam flies by the slit system following a magnetic chicane,
see Fig. \ref{SASE3A2}. } \label{SASE3A}
\end{figure}

\begin{figure}[tb]
\includegraphics[width=1.0\textwidth]{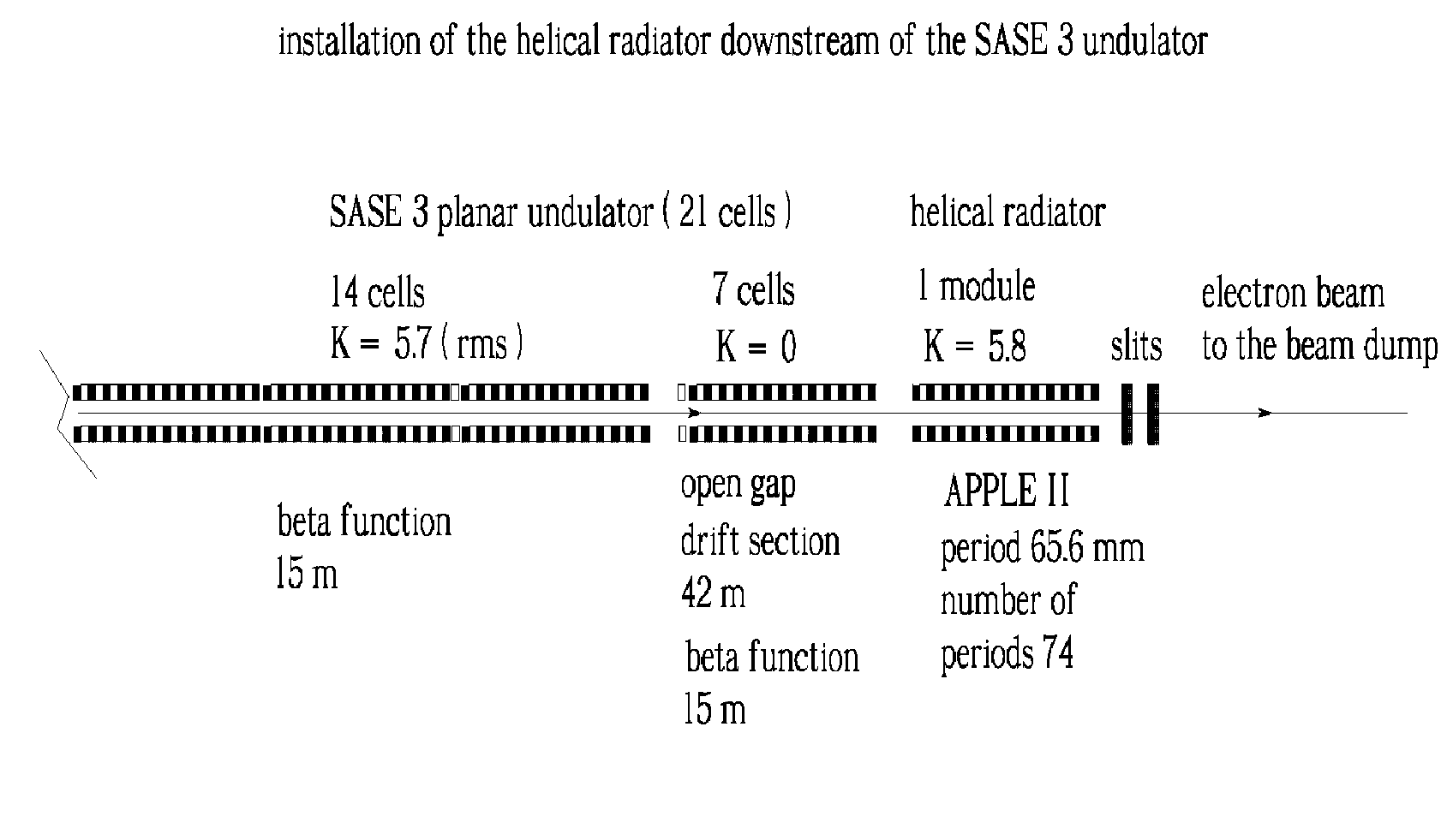}
\caption{The installation of helical radiator and spatial filtering
setup downstream of the SASE3 undulator will allow to produce
high-power, highly circularly-polarized soft X-ray radiation.}
\label{SASE3B}
\end{figure}

\begin{figure}[tb]
\begin{center}
\includegraphics[width=1.0\textwidth]{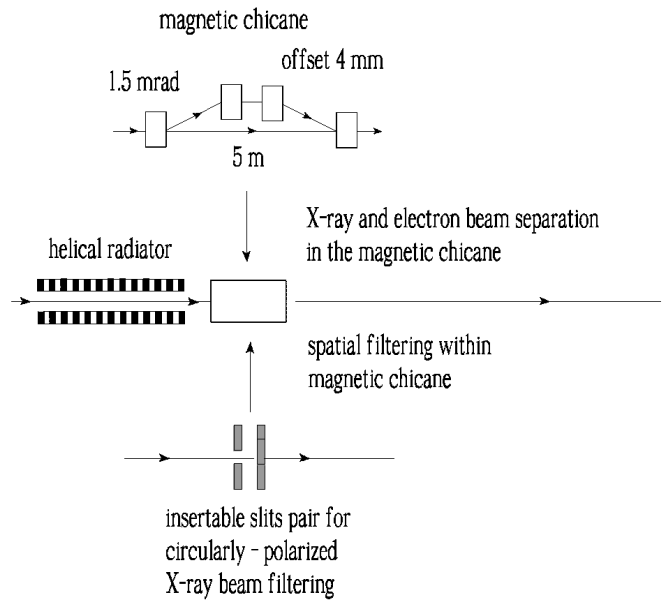}
\end{center}
\caption{The scheme for spatial filtering will make use of a short
magnetic chicane immediately behind the helical radiator, so that
the electron beam can bypass the slits.} \label{SASE3A2}
\end{figure}

\begin{figure}[tb]
\includegraphics[width=1.0\textwidth]{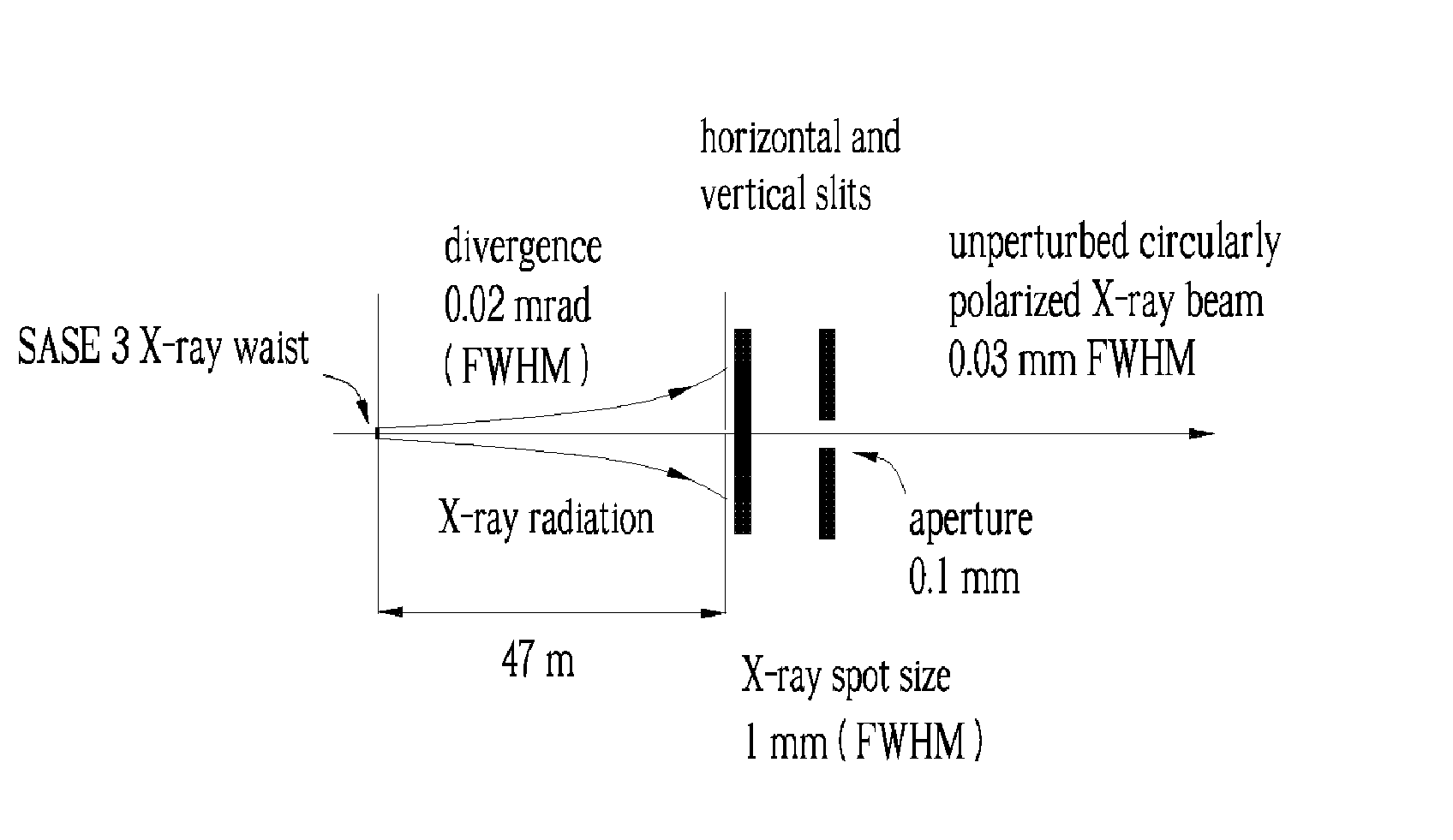}
\caption{Simple method for suppressing the linearly-polarized soft
X-ray radiation from the SASE3 undulator. The linearly-polarized
background can be eliminated by using a special window positioned
downstream of the helical radiator exit. This can be practically
implemented by letting radiation and electron beam through vertical
and horizontal slits positioned 47 m downstream the planar
undulator, where the linearly-polarized radiation pulse is
characterized by a spot size thirty times larger than that of the
circularly-polarized radiation pulse.} \label{SASE3C}
\end{figure}

The principle of our scheme for polarization control at the European
XFEL is illustrated in Fig. \ref{SASE3A}, Fig. \ref{SASE3B},  Fig.
\ref{SASE3A2} and Fig. \ref{SASE3C}.

With reference to Fig. \ref{SASE3A}, the electron beam first goes
through the SASE3 baseline undulator, it produces linearly polarized
SASE radiation, and is modulated both in energy and density.  The
last seven SASE3 undulator modules are left with the gap open, Fig.
\ref{SASE3B}. In this way we provide a total $42$ m - long straight
section for radiation and electron beam transport, corresponding to
the length of the 7 undulator cells. At the end of the straight
section, that is immediately behind the entire (21 cells) SASE3
undulator, we install a 5 m - long APPLE II type undulator module.
While passing through this helical radiator, the microbunched
electron beam produces intense bursts of radiation in any  selected
state of polarization. Subsequently, the circularly polarized
radiation from the APPLE II undulator, the linearly-polarized
radiation from the SASE3 undulator and the electron beam pass
through a spatial filtering station.

A schematic of the spatial filtering station is shown in Fig.
\ref{SASE3A2}. It consists of a magnetic chicane, which allows for
the electron beam to pass by an insertable pair of slits, which is
the main part of the filtering station. Since the slits are
positioned $47$ m downstream of the planar undulator, the
linearly-polarized radiation is characterized by a $0.9$ mm FWHM
spot size, which is about $30$ times larger compared to the
circularly-polarized radiation spot size, amounting to about $0.03$
mm FWHM. At the slit aperture of $0.1$ mm the background radiation
power can therefore be diminished of two orders of magnitude without
perturbing the circularly polarized radiation pulse, Fig.
\ref{SASE3C}. The short chicane allows to decouple the electron beam
trajectory from the radiation. This is required at the European
XFEL, in order to leave complete freedom for the slit design. In
fact, if the electron beam passes through the slits, ionization
losses must be kept to a negligible level, and this poses
constraints on the slit thickness and material. For example, for the
LCLS we previously proposed to use $150~\mu$m-thick Berillium slits
\cite{OURC}. Due to the very high repetition rate and burst
structure of the European XFEL output, such solution is not optimal.
With the use of the chicane instead, the issues of electron beam
transport to the dump and of that of background radiation filtering
are fully decoupled, so that the slit design can fulfill the most
stringent engineering constraints to cope with heat-loading issues.
This idea can straightforwardly be implemented, when needed, at
other facilities as well.

The influence of the propagation of the electron beam through the
drift section on the electron beam microbunching should be accounted
for. In particular, one should account for the fact that the
straight section acts as a dispersive element with momentum
compaction factor $R_{56} \sim 60$ nm at the electron beam energy 14
GeV.  The influence of betatron motion should be further considered.
In fact, the finite angular divergence of the electron beam, which
is linked with betatron function, leads to longitudinal velocity
spread yielding microbunching suppression. In the next Section we
will present a comprehensive study of these effects. We simulated
the evolution of the microbunching along the straight section, and
we concluded that the transport of the microbunched electron beam
through the $42$ m-long straight section does not constitute a
serious problem for the realization of the proposed scheme.

\section{\label{sec:tre} FEL simulations}

\begin{table}
\caption{Parameters for the low-charge mode of operation at the
SASE3 line of the European XFEL used in this paper.}

\begin{small}\begin{tabular}{ l c c}
\hline & ~ Units &  ~ \\ \hline
Undulator period      & mm                  & 68     \\
N periods/module      & -                   & 74   \\
Intersection length   & cm                  & 108  \\
K parameter (rms)     & -                   & 5.7  \\
Wavelength            & nm                  & 1.5   \\
Energy                & GeV                 & 14   \\
Charge                & nC                  & 0.02 \\
Average beta function & m                   & 15   \\
\hline
\end{tabular}\end{small}
\label{tt1}
\end{table}

\begin{figure}[tb]
\includegraphics[width=0.5\textwidth]{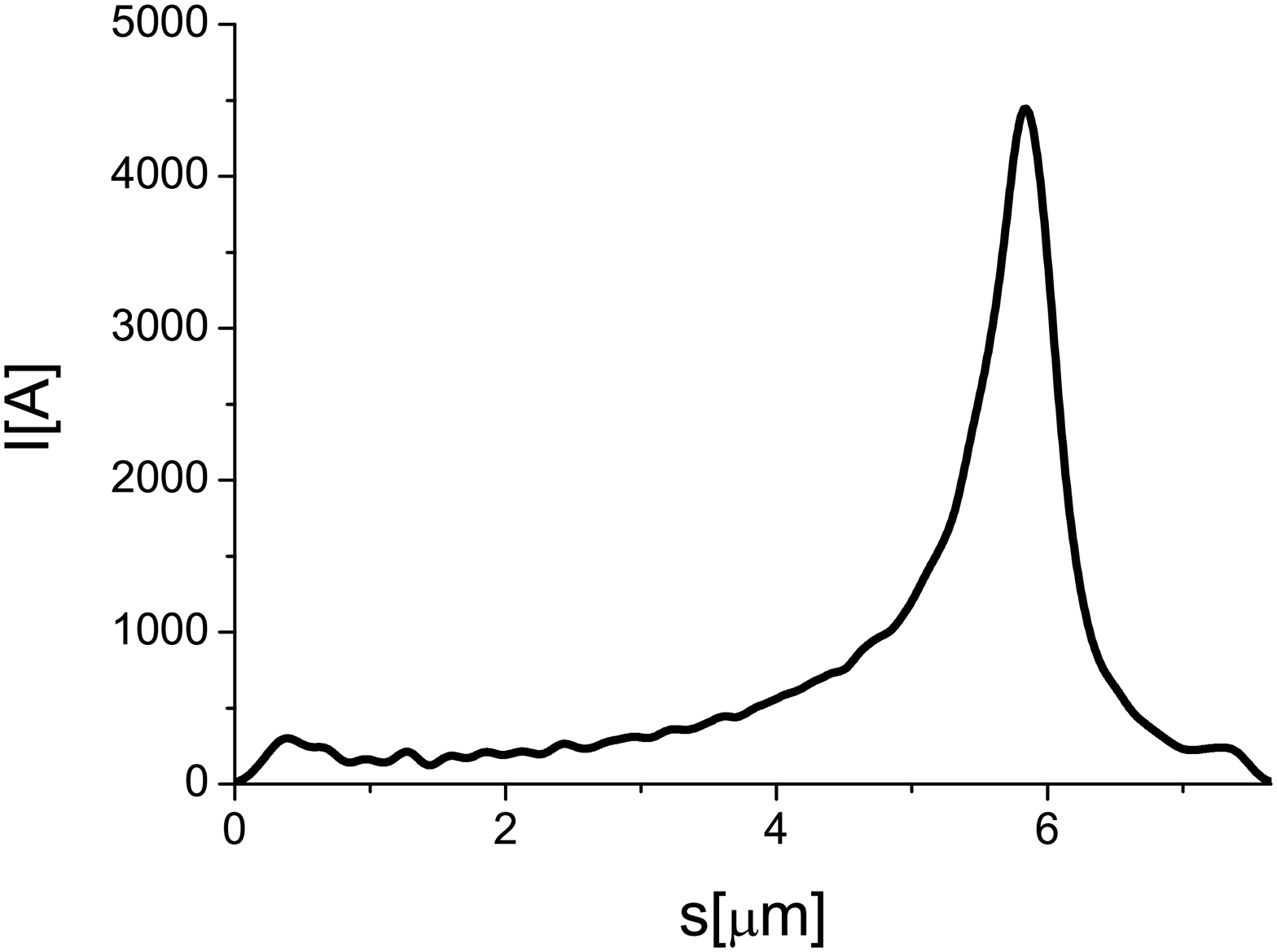}
\includegraphics[width=0.5\textwidth]{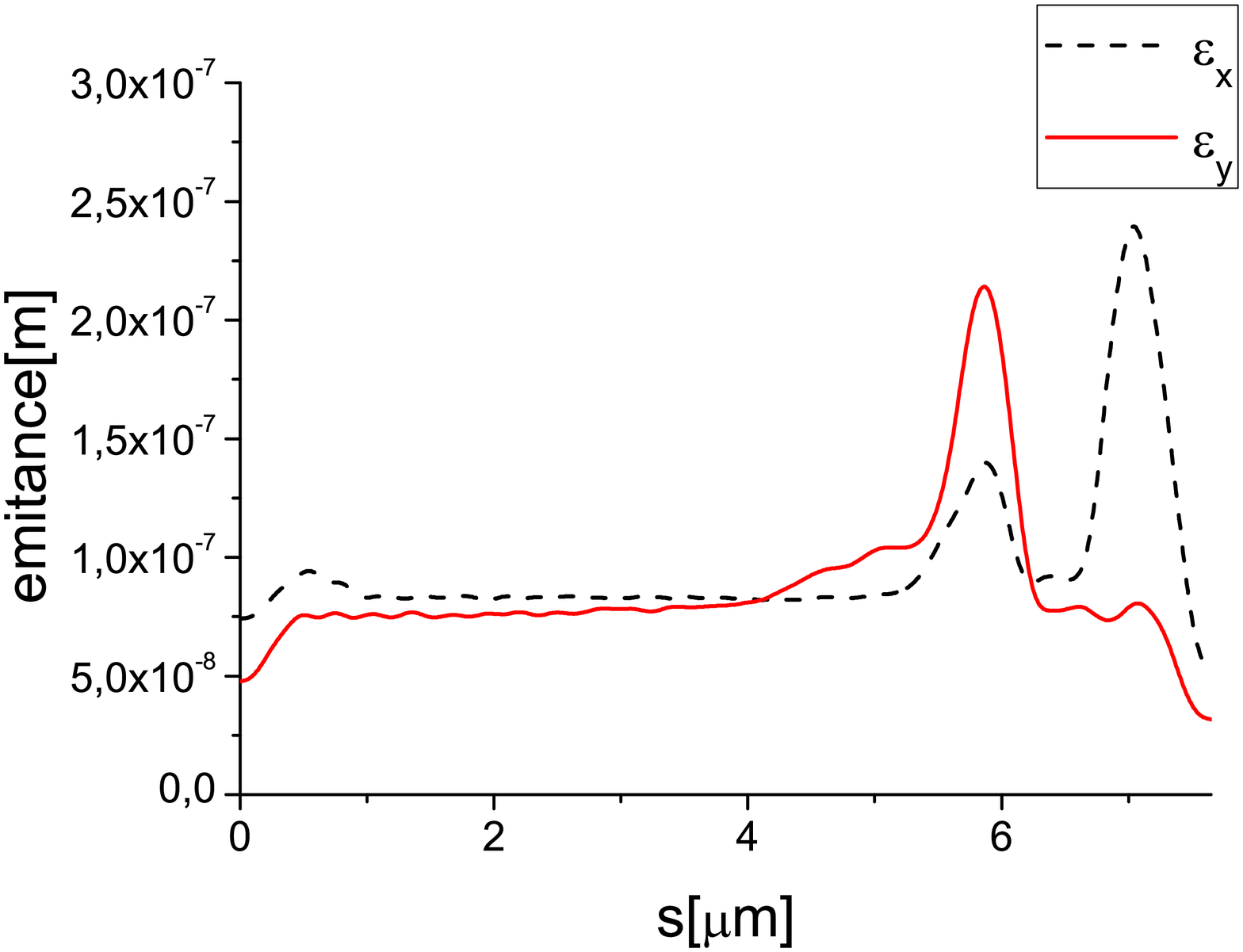}
\includegraphics[width=0.5\textwidth]{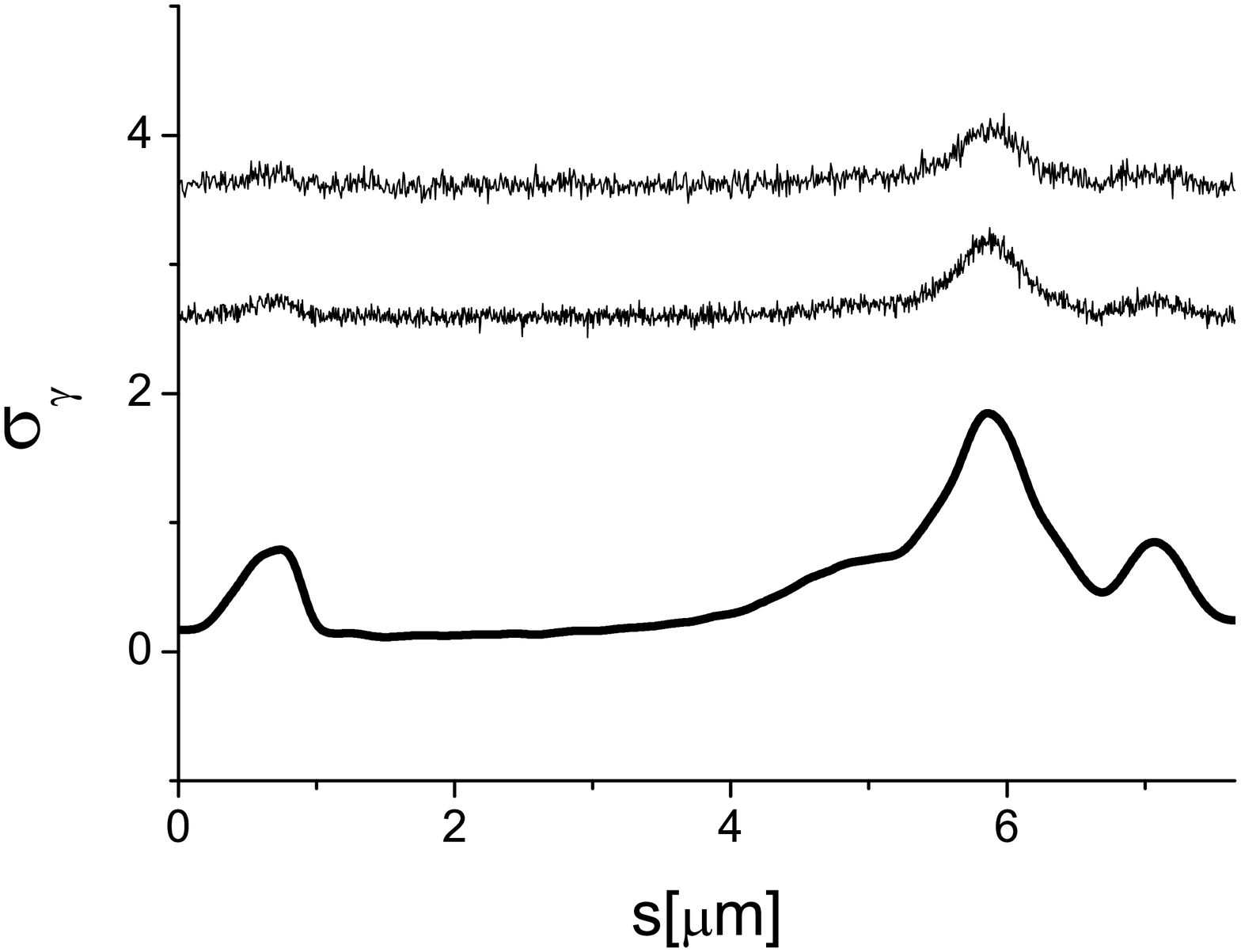}
\includegraphics[width=0.5\textwidth]{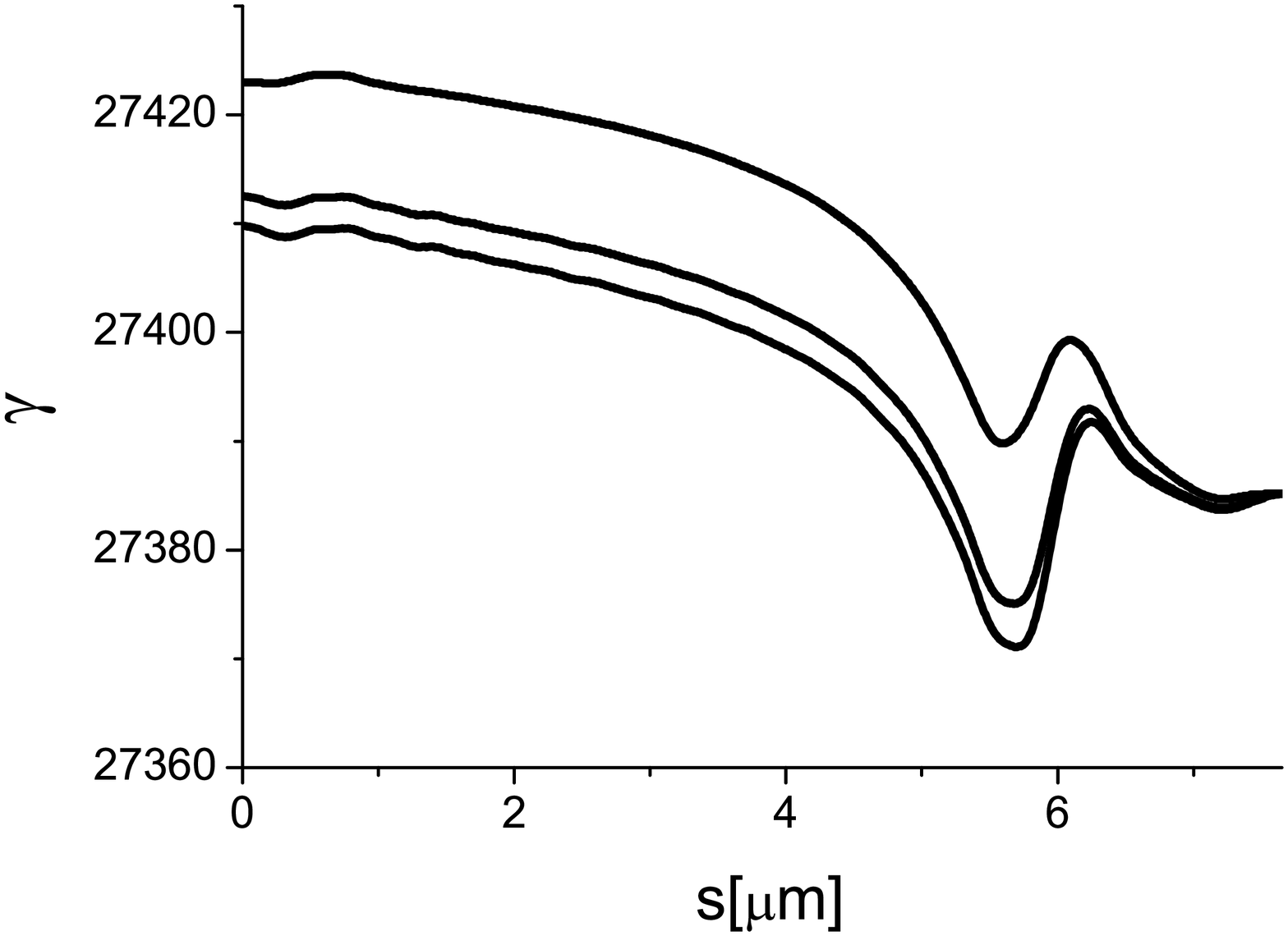}
\caption{The current profile (top left), and the horizontal and
vertical emittances (top right) at the entrance of the SASE1
beamline. Energy spread (bottom left) and energy (bottom right) of
the electrons at different positions down the bemaline: at the
entrance of the SASE1 undulator (lower curve), after SASE1 (middle
curve) and at the entrance of the five working cells of SASE3 (upper
curve).} \label{S2E}
\end{figure}
\begin{figure}[tb]
\includegraphics[width=1.0\textwidth]{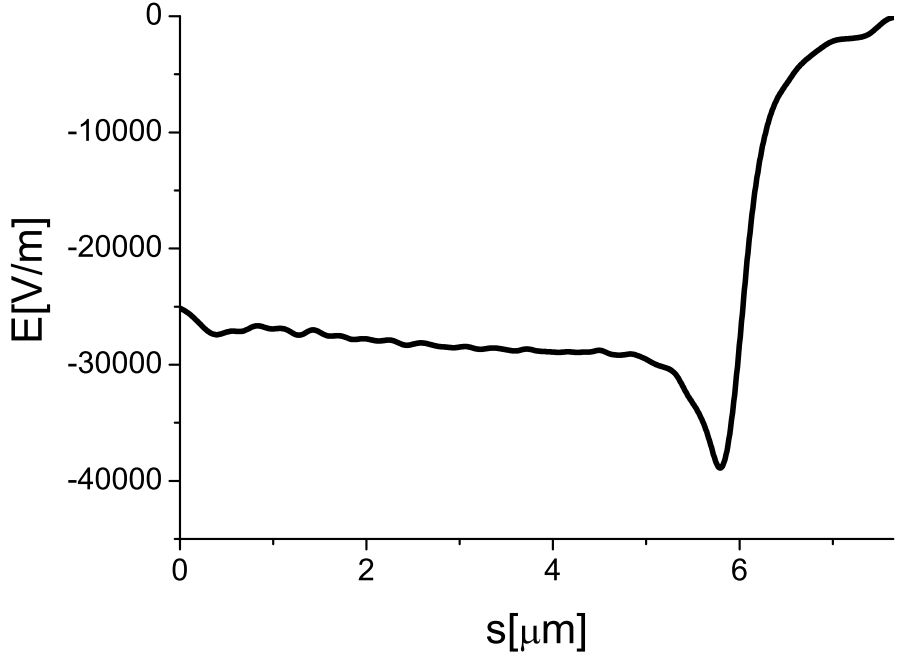}
\caption{The wakefield in SASE1 and in the first 9 cells of SASE3
(detuned).} \label{wake}
\end{figure}

In this Section we report on a feasibility study performed with the
help of the FEL code GENESIS 1.3 \cite{GENE} running on a parallel
machine. We will present a feasibility study for our method of
polarization control at the European XFEL, based on a statistical
analysis consisting of $100$ runs. Parameters used in the
simulations for the low-charge mode of operation ($0.02$ nC) are
presented in Table \ref{tt1}. The choice of the low-charge mode of
operation is motivated by simplicity. The input for GENESIS included
start-to-end simulations of the electron beam characteristics. The
top plots of Fig. \ref{S2E} show the current profile and the
emittances along horizontal and vertical directions at the entrance
of SASE1 \cite{IGOR}. The bottom plots of the same Fig. \ref{S2E}
show the energy and the energy spread at three different locations:
at the entrance of SASE1 following \cite{IGOR} (lower line), at the
exit of SASE1 (middle line) and at the entrance of the $5$ working
cells of SASE3 (upper line). It should be noted here that the output
of the device is optimized when the first 9 SASE3 modules are
detuned from resonance. This means that only the last $5$ SASE3
cells are working. The wake from the undulator vacuum chamber in
Fig. \ref{wake}, also obtained from \cite{IGOR}, and quantum
fluctuations are respectively responsible for the change in energy
and energy spread of the electrons as in the bottom plots of Fig.
\ref{S2E}.

\begin{figure}
\includegraphics[width=0.5\textwidth]{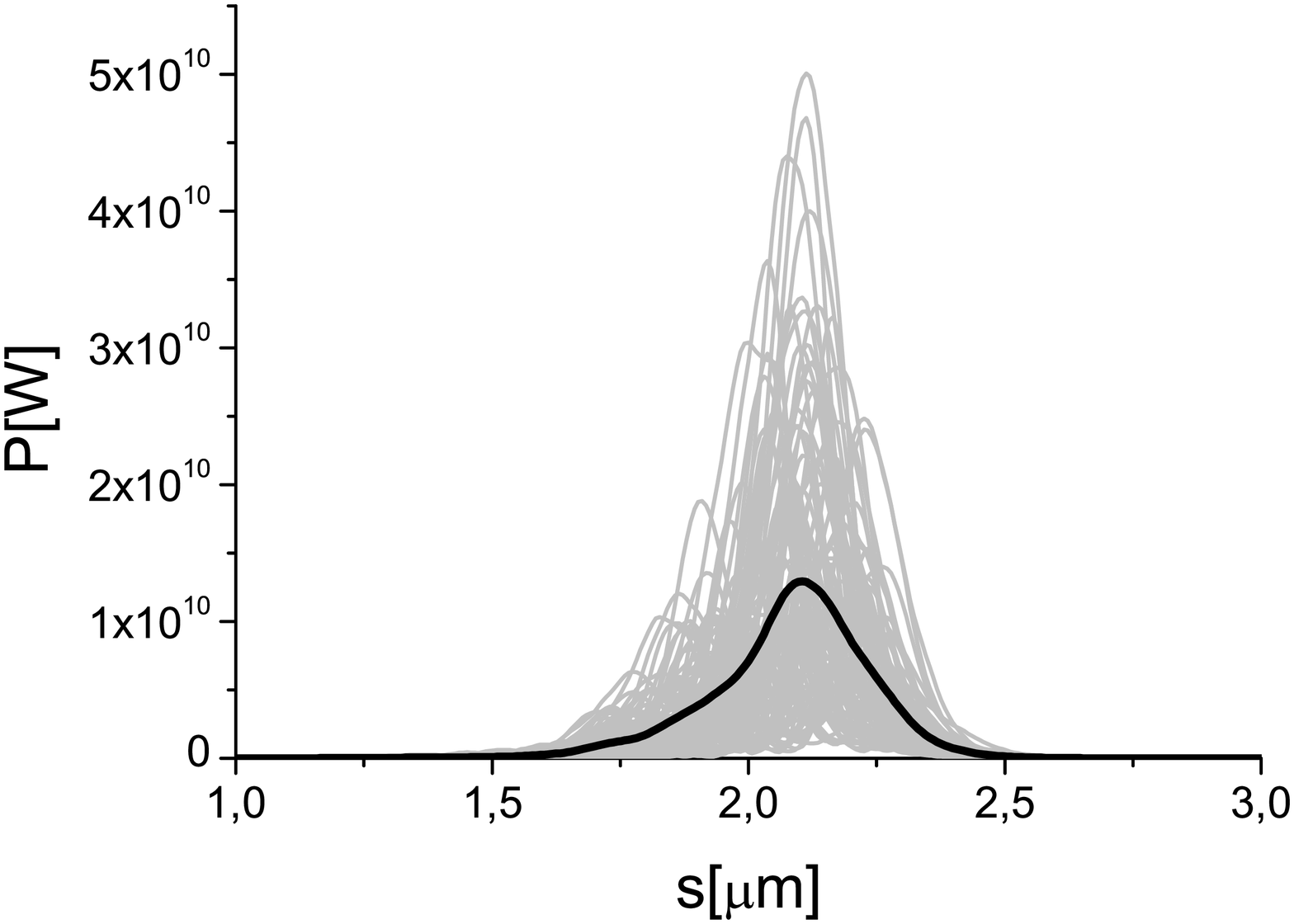}
\includegraphics[width=0.5\textwidth]{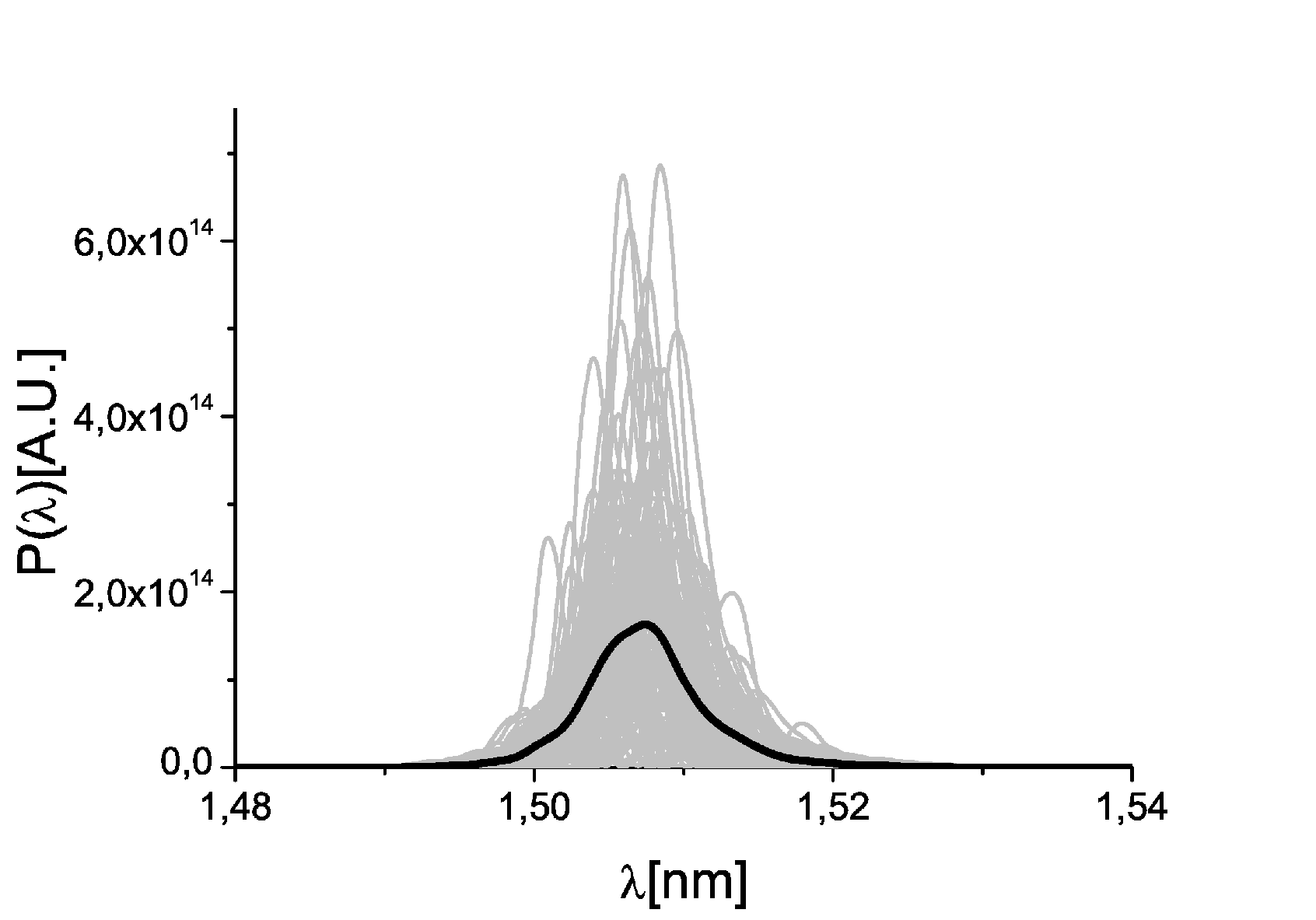}
\caption{Left plot: power distribution after the 5 active SASE3
undulator cells. Right plot: spectrum after the 5 active SASE3
undulator cells. Grey lines refer to single shot realizations, the
black line refers to the average over a hundred realizations. }
\label{SASE}
\end{figure}

\begin{figure}
\includegraphics[width=0.5\textwidth]{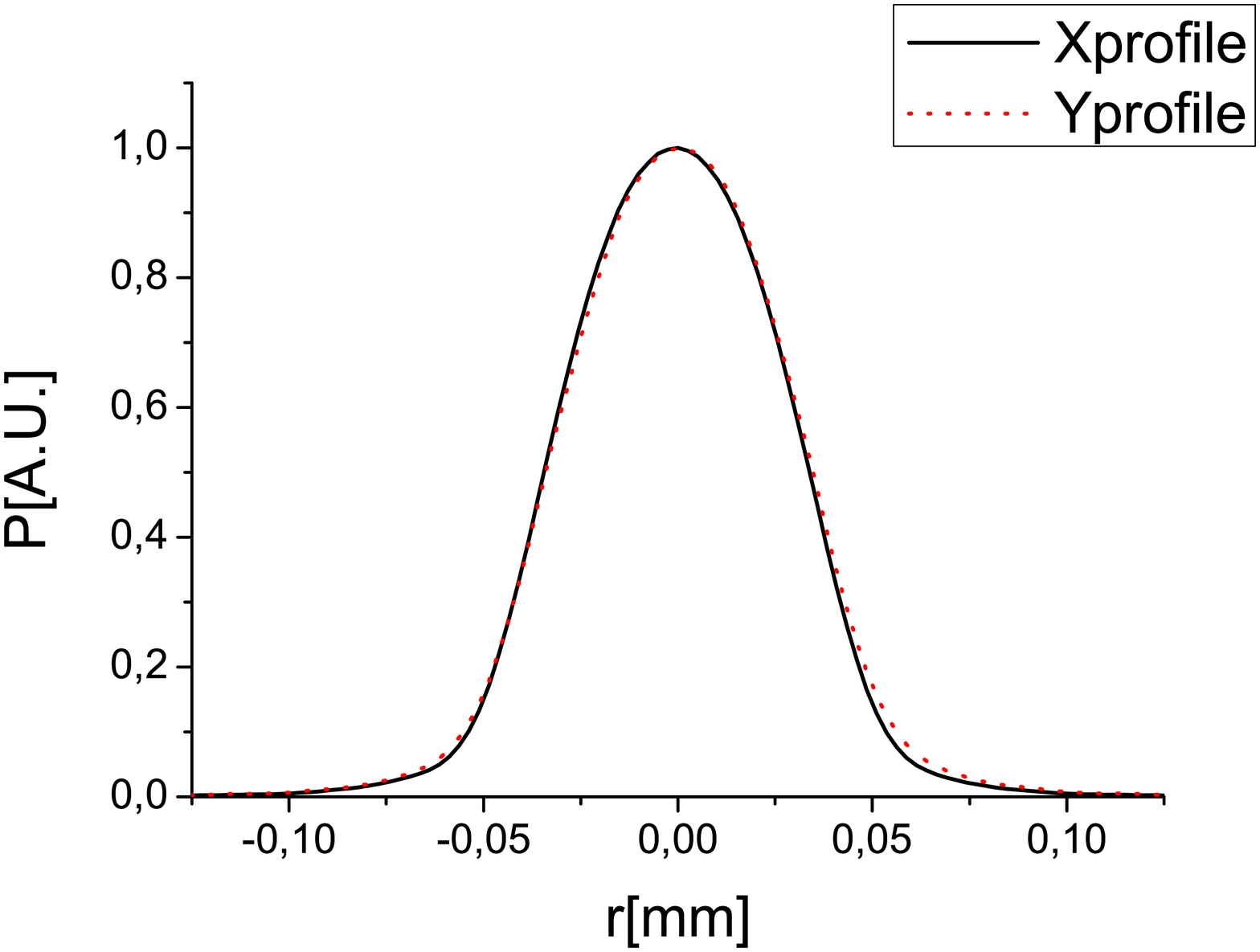}
\includegraphics[width=0.5\textwidth]{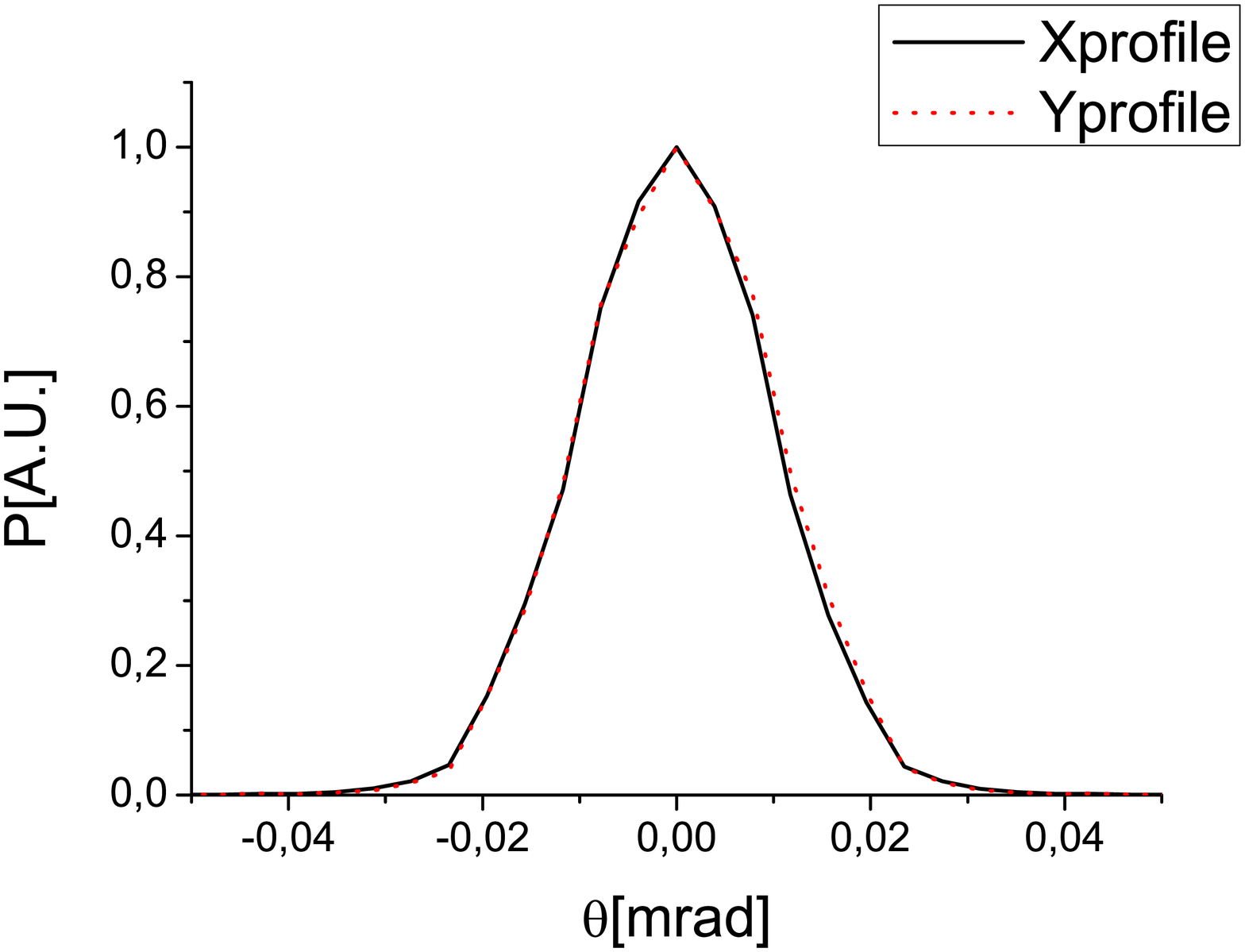}
\caption{Left plot: Transverse plot of the X-ray radiation pulse
energy distribution after the SASE3 undulator (5 active cells).
Right plot: Angular plot of the X-ray radiation pulse energy
distribution after the SASE3 undulator (5 active cells).}
\label{TrAndis}
\end{figure}
First, the baseline SASE undulator output was simulated. The result,
in terms of power and spectrum, is shown in Fig. \ref{SASE}, while
the angular distribution of the radiation is shown in Fig.
\ref{TrAndis}. In order to obtain Fig. \ref{TrAndis}, we first
calculated the intensity distribution along the bunch, so that in
the left plot we present the energy density as a function of the
transverse coordinates $x$ or $y$, as if it was measured by an
integrating photodetector. Finally, a two-dimensional Fourier
transform of each slice of the GENESIS field file was performed.
These Fourier transformed data were squared and summed over the
slice, yielding the angular X-ray radiation pulse energy
distribution. The $x$ and $y$ cuts are shown on the right plot.

\begin{figure}[tb]
\includegraphics[width=1.0\textwidth]{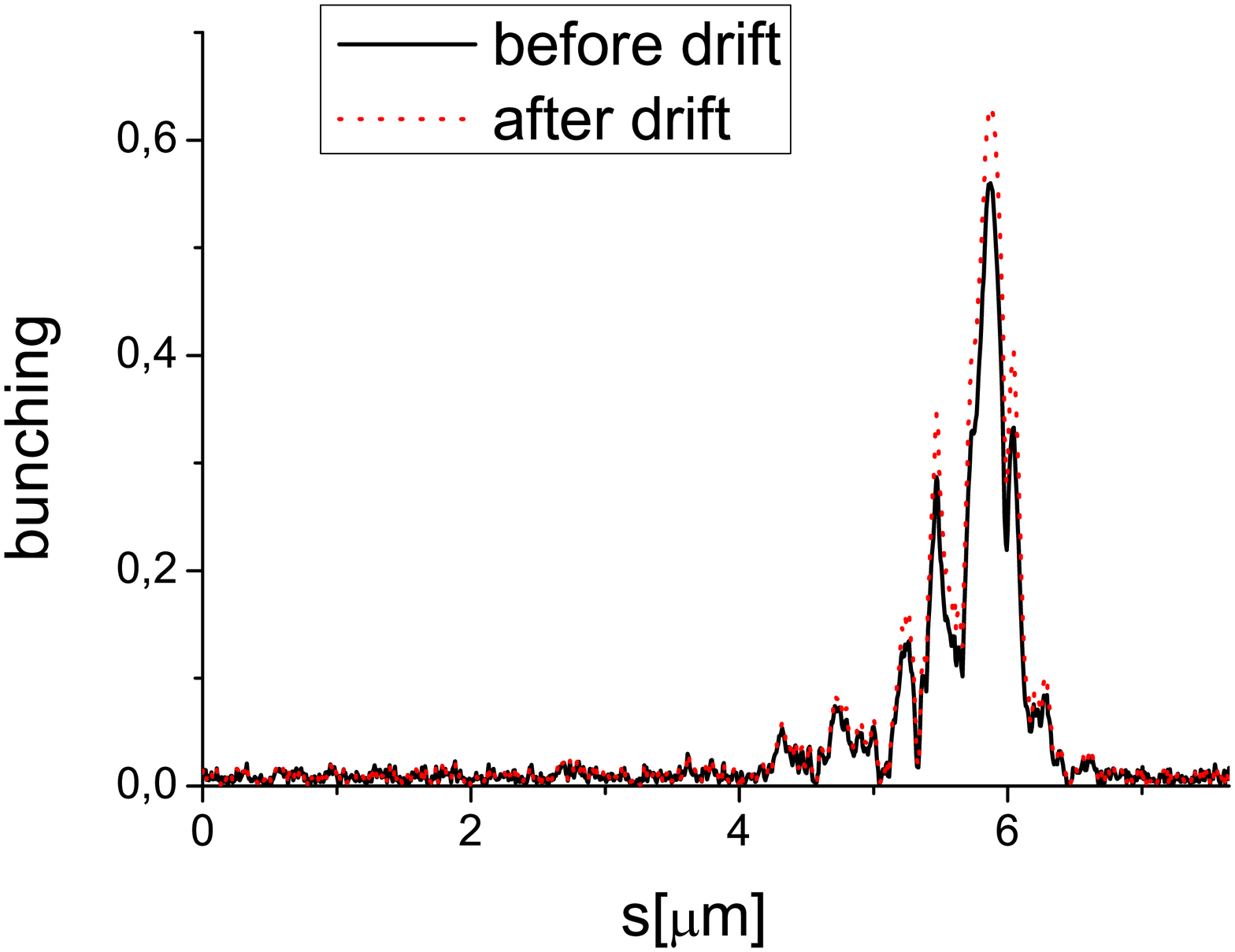}
\includegraphics[width=1.0\textwidth]{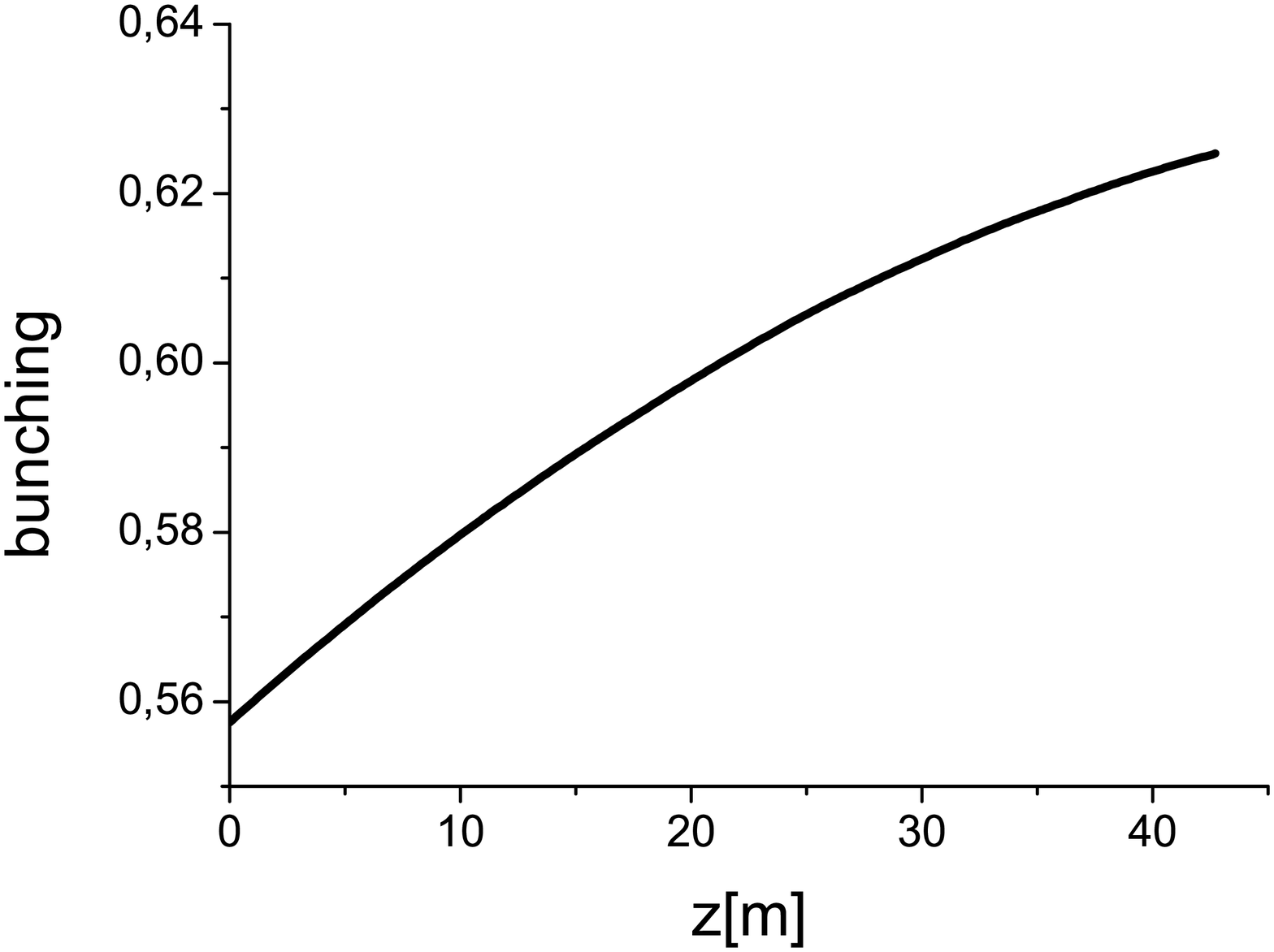}
\caption{Upper plot: comparison between the bunching before and
after the drift for a particular FEL run. Lower plot: evolution of
the bunching at the position of maximal current along the straight
section for the same run.} \label{bunching}
\end{figure}
\begin{figure}[tb]
\includegraphics[width=0.5\textwidth]{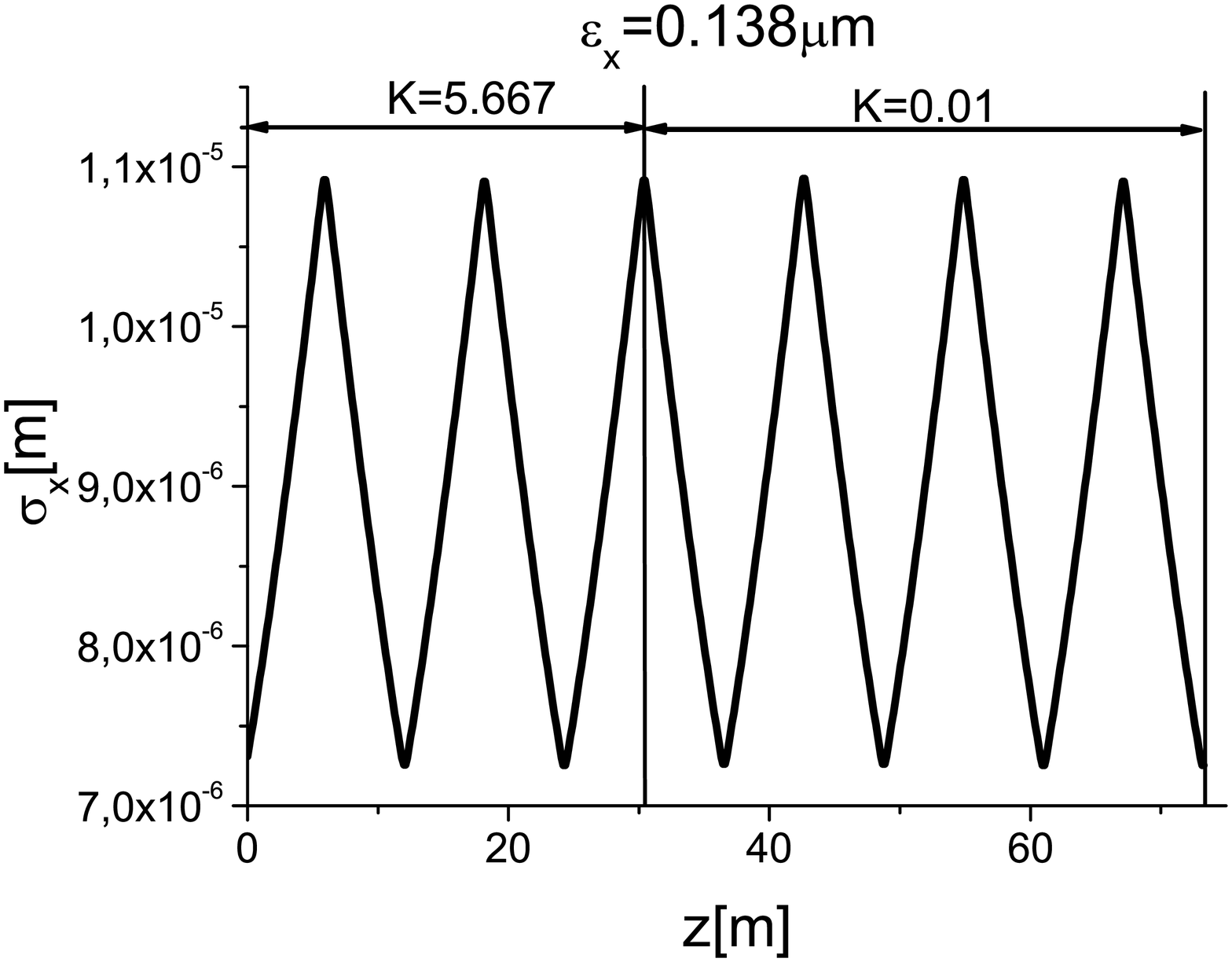}
\includegraphics[width=0.5\textwidth]{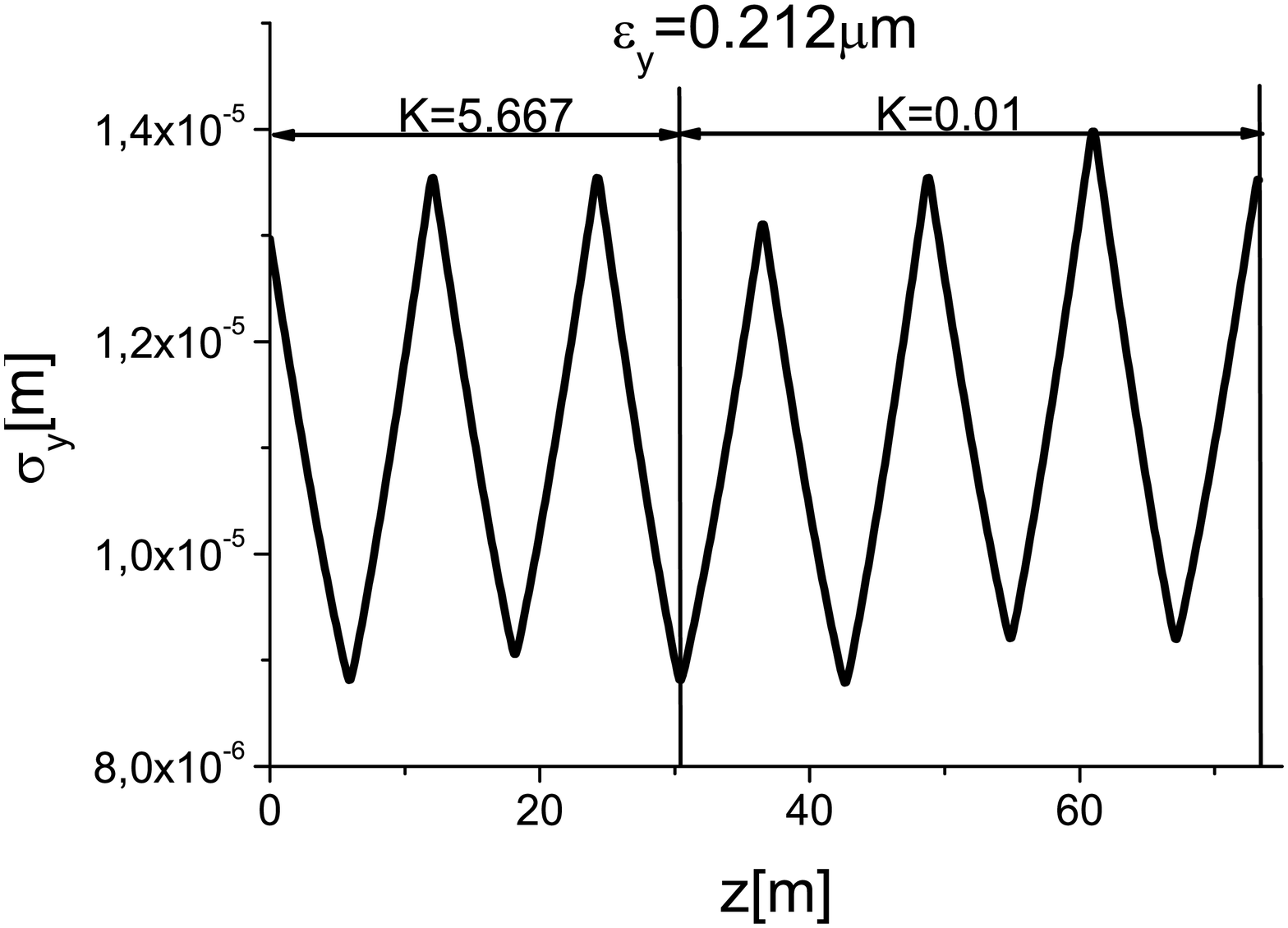}
\caption{Evolution of the rms horizontal (left plot) and vertical
(right plot) beam size as a function of the distance along the setup
calculated through GENESIS. The emittance reported is normalized,
and corresponds to the position within the bunch where the current
is maximal. } \label{SIGXY}
\end{figure}
The GENESIS particle file was downloaded at the exit of the baseline
undulator. For simulating the straight section in GENSIS we used the
same 5-cells undulator structure as for the baseline undulator, but
we changed the undulator parameter to $K = 0.01$ This choice allows
one to have, with negligible differences, the same momentum
compaction factor as in free space. Then the electron beam current
was set to zero, and the undulator focusing was switched off
(although for $K=0.01$ the undulator focusing effects are
negligible). The GENESIS particle file was used as an input for the
propagation of the bunch along the $42$ m-long FODO lattice. GENESIS
automatically accounts for momentum compaction factor and betatron
motion effects on the evolution of the microbunched beam. We tested
the correctness of GENESIS simulations concerning the betatron
motion effects in reference \cite{OUBE}. The bunching before and
after the straight section drift is shown in Fig. \ref{bunching},
upper plot, as a function of the longitudinal coordinate inside the
electron bunch, for a particular run. The evolution of the bunching
at the position of the maximal peak current along the straight
section for the same run is shown in Fig. \ref{bunching}, lower
plot.

The evolution of the rms electron beam size in the horizontal and in
the vertical direction inside the baseline undulator and in the
straight section are shown respectively in Fig. \ref{SIGXY}. These
plots correspond to the position with maximal peak current, and to
normalized slice emittances $\varepsilon_x = 0.138 ~\mu$m and
$\varepsilon_y=0.212~\mu$m. From the average rms beam size and the
normalized emittances corresponding to the slice of maximal peak
current we can conclude that the average betatron function used was
about $15$ m. Inspection of Fig. \ref{SIGXY}, right plot, shows a
little mismatching in the vertical direction $y$, which is not
present in the left plot. This is due to the fact that in baseline
undulator the electron beam was matched accounting for the undulator
focusing properties. However, the mismatch was accounted for,
because we used the particle file at the exit of straight section as
input file for GENESIS simulations of the APPLE II output, meaning
that the particle file was downloaded again at the end of the
propagation through the straight section. Note that each cell begins
with an undulator, and finishes with a quadrupole. Therefore we
downloaded the particle file immediately after the first quadrupole
related with propagation inside the APPLE II undulator. The average
betatron function is again $\beta = 15$ m.  This guarantees correct
propagation along the APPLE II undulator section, which is $5$ m
long. The output files are downloaded immediately after the APPLE II
undulator.

\begin{figure}
\includegraphics[width=0.5\textwidth]{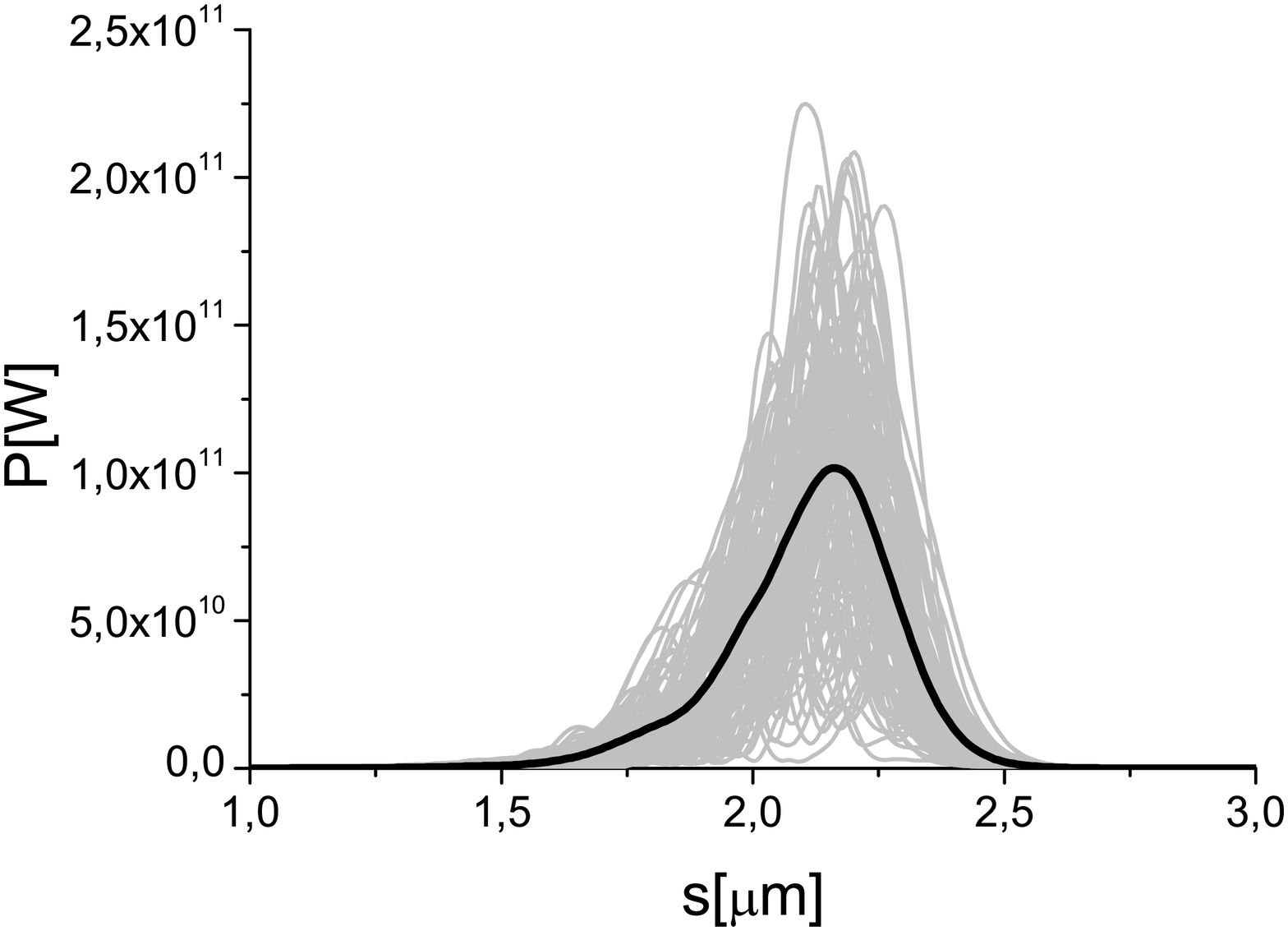}
\includegraphics[width=0.5\textwidth]{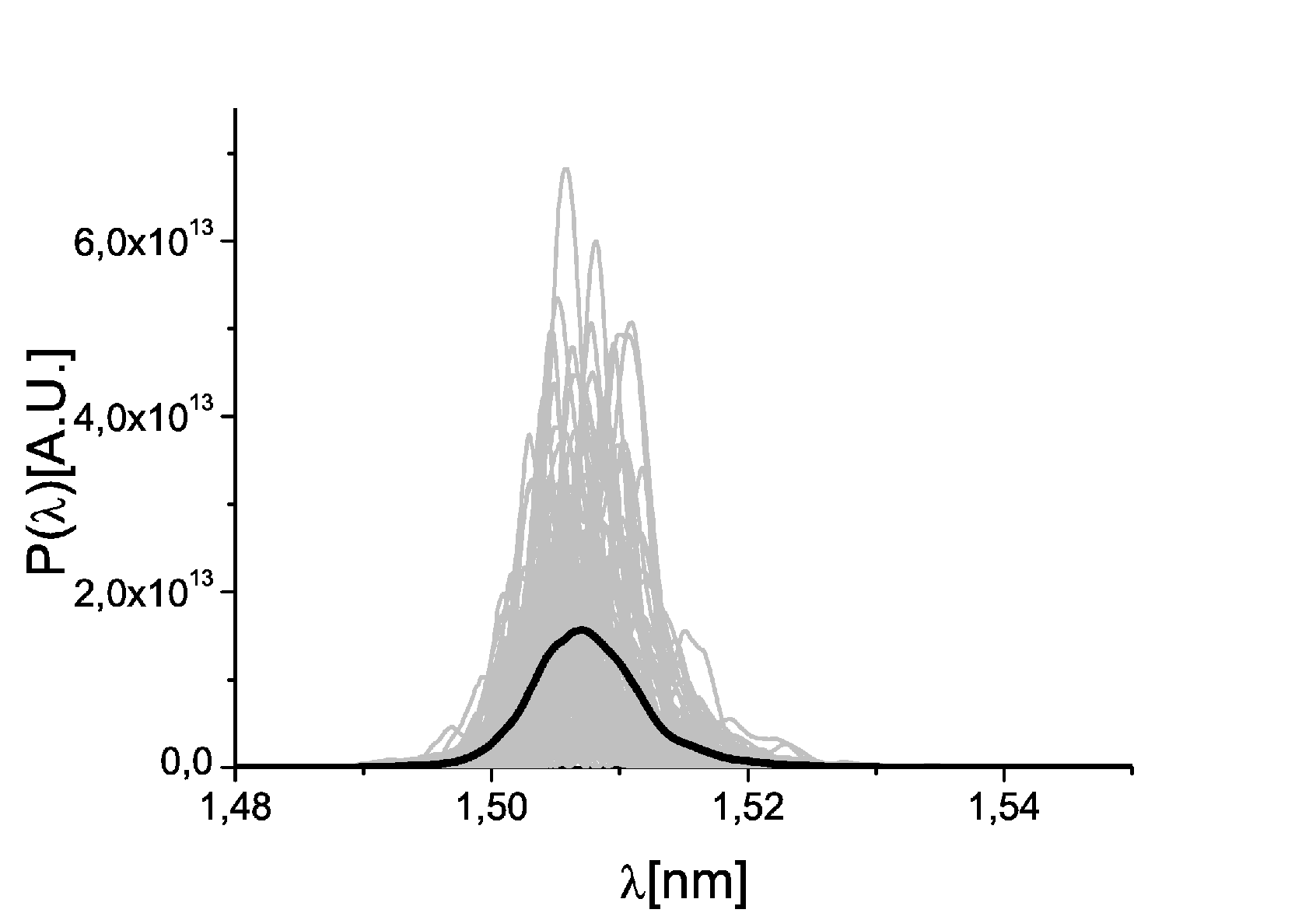}
\caption{Left plot: output power distribution from the APPLE II
undulator.  Right plot: output spectrum. Grey lines refer to single
shot realizations, the black line refers to the average over a
hundred realizations. } \label{output}
\end{figure}

\begin{figure}
\includegraphics[width=0.5\textwidth]{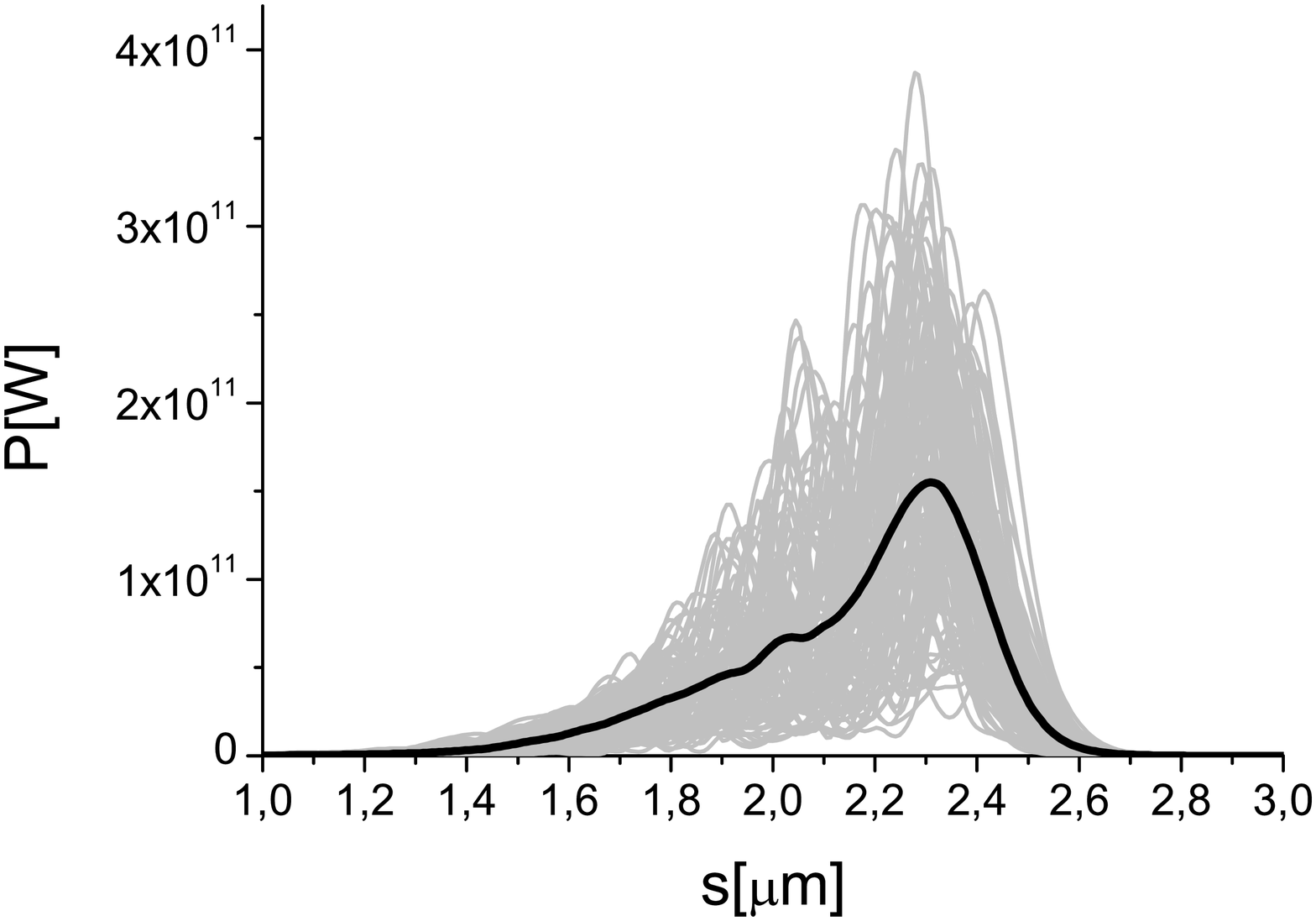}
\includegraphics[width=0.5\textwidth]{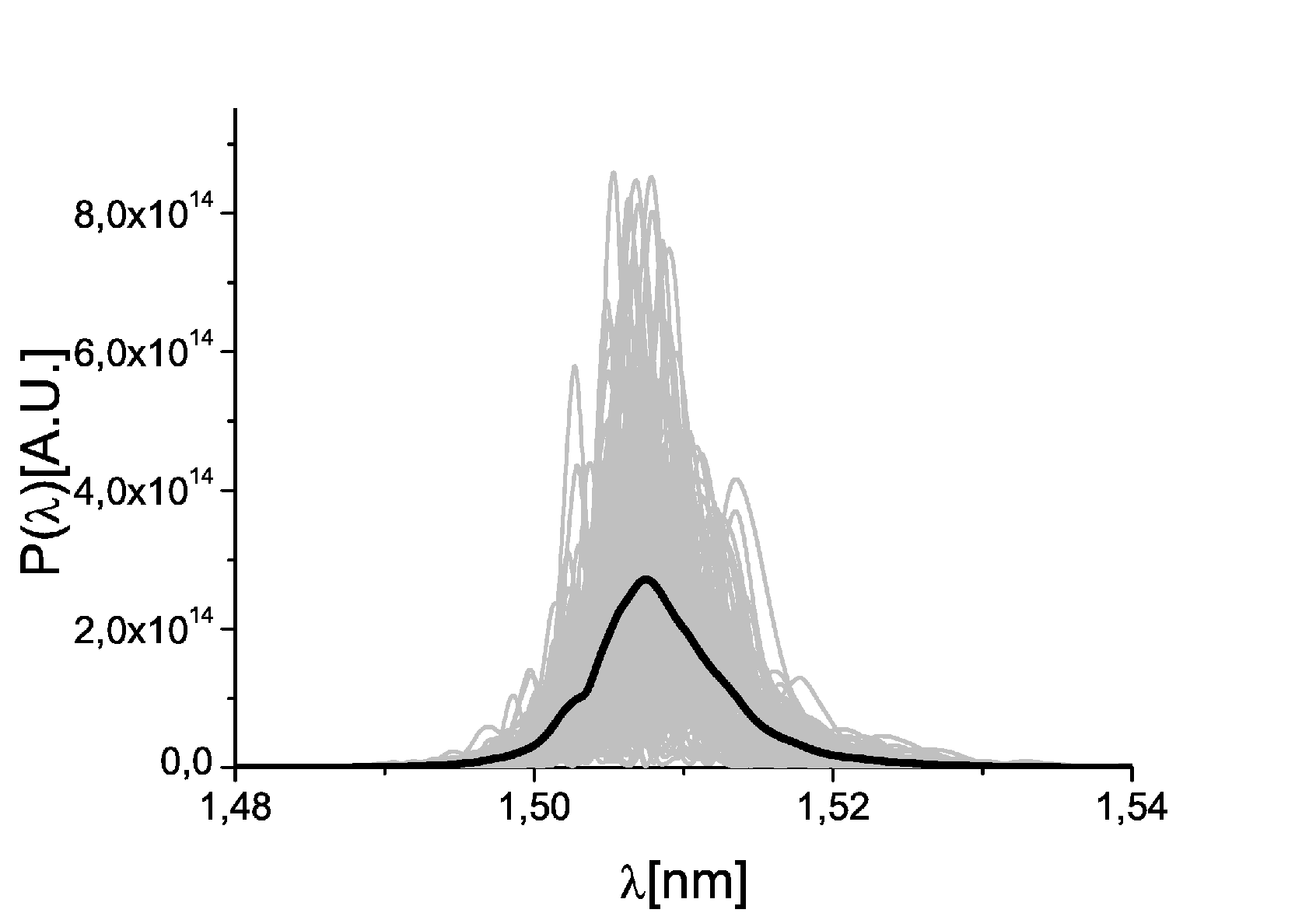}
\caption{Left plot: SASE3 power distribution at saturation (after 7
cells). Right plot: SASE3 spectrum at saturation (after 7 cells).
Grey lines refer to single shot realizations, the black line refers
to the average over a hundred realizations. } \label{SASAT}
\end{figure}
\begin{figure}
\includegraphics[width=0.5\textwidth]{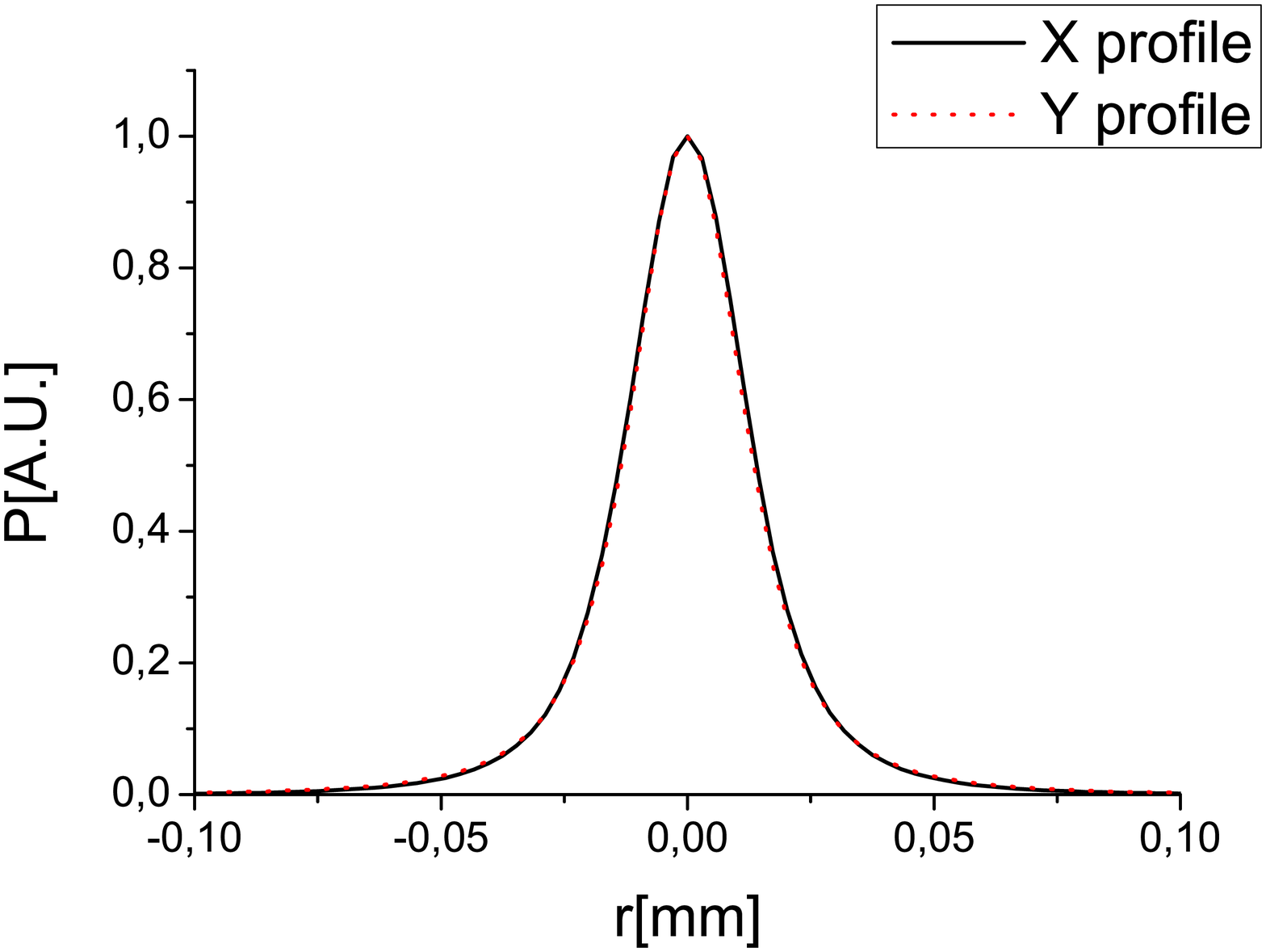}
\includegraphics[width=0.5\textwidth]{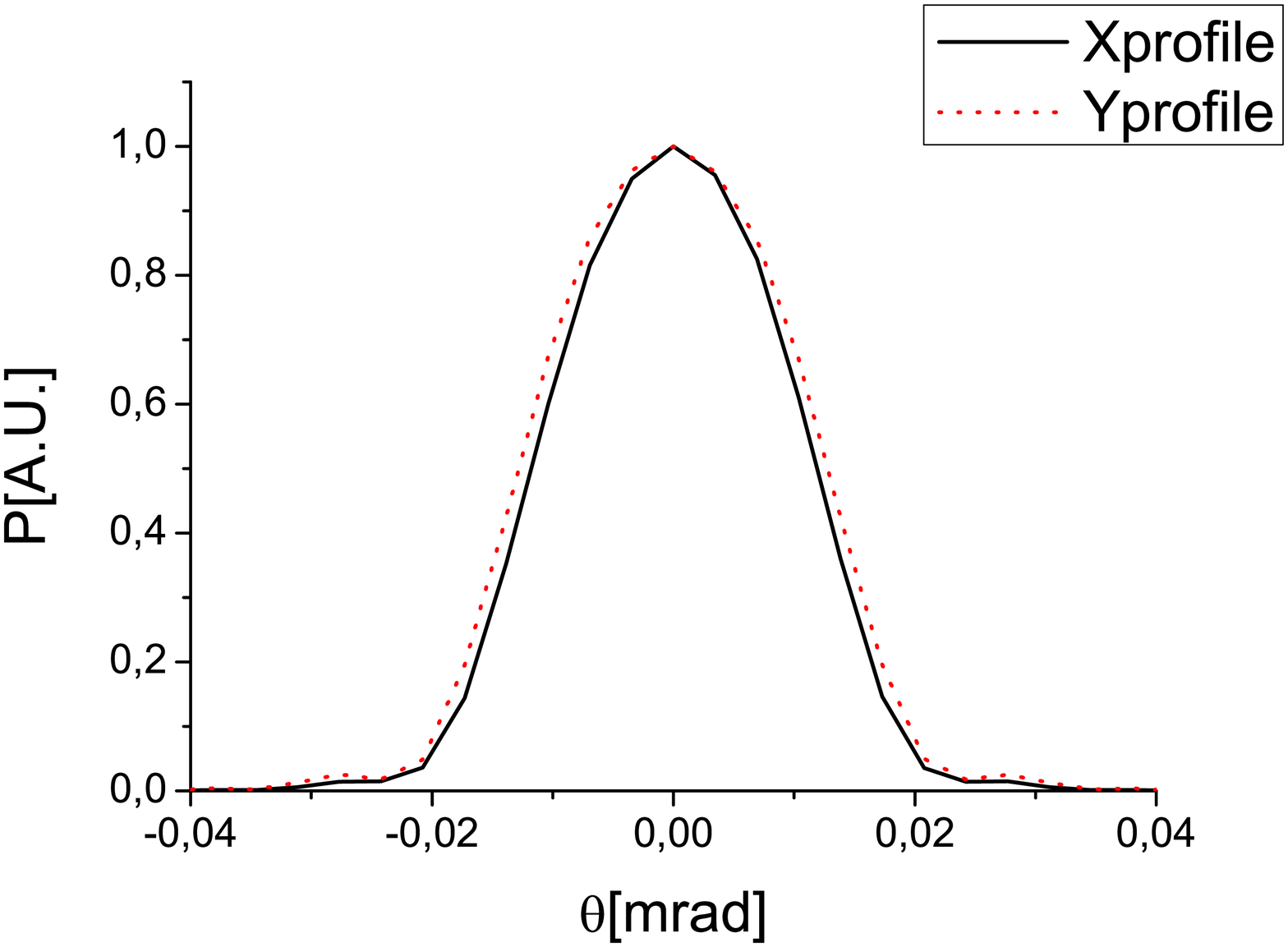}
\caption{Left plot: Transverse plot of the X-ray radiation pulse
energy distribution after the APPLE II undulator (5 cells). Right
plot: Angular plot of the X-ray radiation pulse energy distribution
after after the APPLE II undulator (5 cells).} \label{TrAndis_O}
\end{figure}
The final output from our setup is shown in Fig. \ref{output}, left
plot, in terms of power and in Fig. \ref{output}, right plot, in
terms of spectrum. An output power in the order of $100$ GW is
granted, which is much larger than the SASE level reported in Fig.
\ref{SASE}, and is about the same as the SASE power level at
saturation, which we plot in Fig. \ref{SASAT} for comparison. One
should appreciate the differences compared to the LCLS case
\cite{OURC}, where the APPLE II output is of the same order of
magnitude, but still a factor three smaller than the saturation
level. First, in the SASE3 case we have a much smaller compaction
factor ($R_{56} = 60$ nm compared to $R_{56} = 560$ nm at LCLS), a
factor three smaller emittance, and a factor $1.5$ larger beta
function, while the drift space is only a factor two larger in the
SASE3 setup. As a result, the parameter which controls the
debunching induced by betatron oscillations, $[L
\varepsilon/(\lambda \beta)]^2$ (see \cite{OUBE}) is a factor four
smaller for the SASE3 setup, yielding a smaller effect. The
modulation is almost frozen, and one can optimize the bunching at
the APPLE II entrance to a level very near to the saturation point.
Second, in SASE3 the period is very close to that of the APPLE II
device, as they have practically the same K parameter ($K = 5.66$
rms for the SASE baseline compared to $K = 5.8$ for the APPLE II
undulator). As a result, the APPLE II undulator effectively
interacts with the electron beam. Since a helical undulator can be
considered as a superposition, in the vertical an horizontal planes,
of two planar cells, one can consider a $5$ meter-long APPLE II cell
equivalent to two planar cells, which help in the building up of
extra-bunching, and in the overall output increase.

The transverse distribution of the radiation is shown in Fig.
\ref{TrAndis_O} in terms of transverse coordinates (left plot) and
angles (right plot). From the analysis of Fig. \ref{TrAndis} one
finds an angular size in the order of $20~\mu$rad FWHM. As a result,
after $47$ m propagation (including the drift an the APPLE II
undulator), the transverse size of the SASE radiation is about $0.9$
mm FWHM, to be compared with the APPLE II radiation spot size, which
is about $30~\mu$m. The SASE radiation spot size is, therefore,
about $30$ times larger than the APPLE II radiation spot size.
Moreover, the ratio between circularly and linearly polarized peak
power is about $10$. A slit system letting through the FWHM of the
APPLE II radiation would let pass a relative background radiation
contribution (linearized polarized) of about $10^{-1} \cdot
(30/900)^2 \sim 10^{-4}$, which is practically negligible.

\section{\label{sec:conc} Conclusions}

The European XFEL does not offer the possibility of polarization
control: the output radiation is simply linearly polarized. However,
the production of light with variable polarization, and in
particular of circularly polarized radiation is of most interest for
many experiments, prevalently in the soft X-ray region. In this
paper we demonstrated that such polarization control can be achieved
with the help of an almost trivial setup, composed by three
components: an APPLE II type undulator module, which is meant to be
installed immediately behind the SASE3 undulator line, a short
magnetic chicane, to be installed  immediately behind the APPLE II
undulator module, and an insertable slits pair, which is installed
in the transverse offset created by the magnetic chicane, and serves
for X-ray spatial filtering. The setup can be straightforwardly
installed behind the SASE3 undulator and is safe, in the sense that
it guarantees the baseline mode of operation. The proposed
polarization control setup takes almost no cost and time to be
implemented at the European XFEL.

\section{Acknowledgements}

We are grateful to Massimo Altarelli, Reinhard Brinkmann, Serguei
Molodtsov and Edgar Weckert for their support and their interest
during the compilation of this work.

\end{document}